\documentclass[12pt,a4paper]{article}

\usepackage[T1]{fontenc}
\usepackage[utf8]{inputenc}
\usepackage[nosort]{cite}
\usepackage{color}
\usepackage[bulletsep]{collref}
\usepackage{graphicx}
\usepackage{bbm}
\usepackage{amsmath}
\usepackage{amssymb}
\usepackage{subfig}
\usepackage{verbatim}
\usepackage{multirow}
\usepackage{tikz}
\usepackage{amsthm}
\usepackage{fancyhdr}
\usepackage{wrapfig}
\usepackage{hyperref}
\usepackage{dsfont}%for identity operator \idop
\usepackage[extdef]{delimset} % for brackets

\makeatletter \@addtoreset{equation}{section} \makeatother

\makeatletter
\let\old@startsection=\@startsection
\let\oldl@section=\l@section
\renewcommand{\@startsection}[6]{\old@startsection{#1}{#2}{#3}{#4}{#5}{#6\mathversion{bold}}}
\renewcommand{\l@section}[2]{\oldl@section{\mathversion{bold}#1}{#2}}
\makeatother

\makeatletter
\let\old@makecaption=\@makecaption
\def\@makecaption{\small\old@makecaption}
\makeatother

\let\oldPhi=\Phi
\let\oldPsi=\Psi
\let\oldGamma=\Gamma
\let\oldDelta=\Delta
\let\oldSigma=\Sigma
\let\oldTheta=\Theta
\let\oldPi=\Pi
\let\oldUpsilon=\Upsilon
\renewcommand{\Phi}{\mathnormal{\oldPhi}}
\renewcommand{\Psi}{\mathnormal{\oldPsi}}
\renewcommand{\Gamma}{\mathnormal{\oldGamma}}
\renewcommand{\Sigma}{\mathnormal{\oldSigma}}
\renewcommand{\Delta}{\mathnormal{\oldDelta}}
\renewcommand{\Theta}{\mathnormal{\oldTheta}}
\renewcommand{\Pi}{\mathnormal{\oldPi}}
\renewcommand{\Upsilon}{\mathnormal{\oldUpsilon}}

\newcommand{\superN}{\mathcal{N}}

\newcommand{\Lagr}{\mathcal{L}}

\newcommand{\tr}{\mathop{\mathrm{tr}}}

\newcommand{\fd}{{\text{extra}}}

\makeatletter
\newlength{\apb@width}
\newcommand{\autoparbox}[2][c]{\settowidth{\apb@width}{#2}\parbox[#1]{\apb@width}{#2}}
\newcommand{\includegraphicsbox}[2][]{\autoparbox{\includegraphics[#1]{#2}}}
\makeatother
\ifx\genfrac\sdflkaj

\else

\fi
\newcommand{\sfrac}[2]{{\textstyle\frac{#1}{#2}}}
\newcommand{\half}{\sfrac{1}{2}}
\newcommand{\ihalf}{\sfrac{i}{2}}
\newcommand{\quarter}{\sfrac{1}{4}}

\newcommand{\alg}[1]{\mathfrak{#1}}
\newcommand{\grp}[1]{\mathrm{#1}}

\newcommand{\primed}[1]{{ #1'}}

\makeatletter
\def\mr@ignsp#1 {\ifx\:#1\@empty\else #1\expandafter\mr@ignsp\fi}%
\newcommand{\multiref}[1]{\begingroup%\let\protect\string%
\xdef\mr@no@sparg{\expandafter\mr@ignsp#1 \: }%
\def\mr@comma{}%
\@for\mr@refs:=\mr@no@sparg\do{\mr@comma\def\mr@comma{,}\ref{\mr@refs}}%
\endgroup}
\makeatother

\newcommand{\hypref}[2]{\ifx\href\asklfhas #2\else\href{#1}{#2}\fi}
\newcommand{\Secref}[1]{Section~\multiref{#1}}
\newcommand{\secref}[1]{section~\multiref{#1}}
\newcommand{\Appref}[1]{Appendix~\multiref{#1}}

\newcommand{\Tabref}[1]{Table~\multiref{#1}}

\newcommand{\Figref}[1]{Figure~\multiref{#1}}

\renewcommand{\eqref}[1]{(\multiref{#1})}

\ifx\href\asklfhas\newcommand{\href}[2]{#2}\fi

\newcommand{\eps}{\varepsilon}

\newcommand{\be}{\begin{eqnarray}}
\newcommand{\ee}{\end{eqnarray}}

\newcommand{\gen}[1]{\mathrm{#1}}

\newcommand{\phione}{Z}
\newcommand{\phitwo}{X}
\newcommand{\m}{z}

\usepackage{color}
\begin{document}

\thispagestyle{empty}

\begin{flushright}\footnotesize
%\texttt{arXiv:xxxx.xxxx}\\
\texttt{HU-EP-20/21}%
\end{flushright}
\vspace{1cm}

\begin{center}%
{\LARGE\textbf{\mathversion{bold}%
Massive Fishnets
}\par}
\medskip

\vspace{1.2cm}

 \textsc{Florian Loebbert, Julian Miczajka} \vspace{8mm} \\
\textit{%
Institut f\"{u}r Physik, Humboldt-Universit\"{a}t zu Berlin, \\
Zum Gro{\ss}en Windkanal 6, 12489 Berlin, Germany
} \\

\texttt{\\ \{loebbert,miczajka\}@physik.hu-berlin.de}

%%%%%%%%
\par\vspace{10mm}

\textbf{Abstract} \vspace{5mm}

\begin{minipage}{12.85cm}
Recently, infinite families of massive Feynman integrals were found to feature an unexpected Yangian symmetry. In the massless case,
similar integrability properties are understood via the interpretation of individual Feynman integrals as correlators in the massless fishnet theory introduced by G{\"u}rdo{\u{g}}an and Kazakov. Here we seek for an analogous interpretation of the integrability of massive Feynman integrals.
We contrast two approaches to define simple massive quantum field theories in four dimensions. First, we 
discuss spontaneous symmetry breaking in the massless bi-scalar fishnet theory. 
We then propose an alternative route to a massive fishnet theory by taking a double-scaling limit of $\superN=4$ SYM theory on the Coulomb branch. Both approaches lead to a massive extension of the massless fishnet theory, differing in how masses enter into the propagators. In the latter theory, planar off-shell amplitudes are in one-to-one correspondence with precisely those massive Feynman integrals that were shown to be invariant under the Yangian. This suggests a re-investigation of Coulomb branch $\superN=4$ SYM theory with regard to integrability. Finally, we demonstrate that in the case of spontaneous symmetry breaking, the original conformal symmetry leads to soft theorems for scattering amplitudes in the broken phase. 

\end{minipage}
\end{center}
\newpage

\tableofcontents
\bigskip
\hrule

%%%%%%%%%%%%%%%%%%%%%%%%%%%%%%%%%%%%%%%%%%%%%%%%%%%%%%%%%%%%%%%%%%%%%%%%%%%
%%%%%%%%%%%%%%%%%%%%%%%%%%%%%%%%%%%%%%%%%%%%%%%%%%%%%%%%%%%%%%%%%%%%%%%%%%%%
%%%%%%%%%%%%%%%%%%%%%%%%%%%%%%%%%%%%%%%%%%%%%%%%%%%%%%%%%%%%%%%%%%%%%%%%%%%
\section{Introduction}

The framework of quantum field theory (QFT) is arguably one of the most advanced tools of theoretical physics. At the same time computations quickly become untractable and are limited to a perturbative regime. This underlines the demand for toy models that are sufficiently simple to progress computations, while still featuring enough complexity to study generic properties of quantum field theory.
More than 20 years ago the celebrated AdS/CFT duality opened a new door to unveil such toy models of QFT --- first of all $\superN=4$ super Yang--Mills (SYM) theory \cite{Maldacena:1997re}.
In these models, simplicity goes hand in hand with a rich spectrum of symmetries. 
In particular, integrability has proven extremely powerful for computations in various instances of the AdS/CFT correspondence \cite{Beisert:2010jr,Bombardelli:2016rwb}. The appearence of integrable structures in massless quantum field theories in four dimensions is strongly intertwined with the duality to two-dimensional worldsheet string models. Their self-duality results in two instances of conformal symmetry which in the most prominent cases close into a Yangian algebra \cite{Dolan:2003uh,Drummond:2009fd}. This infinite-dimensional symmetry is highly constraining and provides the algebraic underpinning of rational integrable models \cite{Bernard:1992ya,MacKay:2004tc,Torrielli:2011gg,Loebbert:2016cdm}.
\bigskip

The Yangian symmetry is inherited by large classes of fishnet Feynman integrals via their interpretation as scattering amplitudes or correlation functions in the so-called fishnet theories~\cite{Gurdogan:2015csr,Caetano:2016ydc,Chicherin:2017cns,Chicherin:2017frs}.
 The massless fishnet theories have been defined as particular double-scaling limits of gamma-deformed $\superN=4$ SYM theory and represent inspiring toy models of quantum field theory in four dimensions, see \cite{
 Chicherin:2017cns,
 Chicherin:2017frs,
 Sieg:2016vap,
Mamroud:2017uyz,
Gromov:2017cja,
Grabner:2017pgm,
Kazakov:2018qbr,
Basso:2018agi,
Gromov:2018hut,
Derkachov:2018rot,
Korchemsky:2018hnb,
Ipsen:2018fmu,
Basso:2018cvy,
Kazakov:2018gcy,
deMelloKoch:2019ywq,
Gromov:2019aku,
Pittelli:2019ceq,
Gromov:2019bsj,
Chowdhury:2019hns,
Karananas:2019fox,
Adamo:2019lor,
Gromov:2019jfh,
Basso:2019xay,
Derkachov:2019tzo,
Levkovich-Maslyuk:2020rlp,
Karananas:2020lpv,
Wu:2020nis,
Derkachov:2020zvv,
1813022} for progress in various directions. They have a simple field content, are believed to be integrable and require no supersymmetry. The price to pay for these features is the lack of unitary. While non-unitarity may be considered a significant drawback, it is compensated by a one-to-one correspondence between correlation functions (or scattering amplitudes) and Feynman integrals. This correspondence allows to translate the symmetries of correlators directly to the fundamental building blocks of generic quantum field theories:
 \begin{equation}
\mathcal{N}=4 \text{ SYM} 
\quad\to\quad
\gamma\text{-deformation}
\quad\to\quad
\text{fishnet theory} 
\quad\to\quad
\text{Feynman graphs}
\label{eq:Sequence}
\end{equation}
Hence, the integrability features of the AdS/CFT duality find their way to phenomenologically interesting quantities and can be used to fix certain Feynman integrals completely~\cite{Loebbert:2019vcj}.
\bigskip

In particular with regard to phenomenology, the interest in Feynman integrals is not limited to the case of massless propagators. This raises the question of whether the above sequence of steps can be generalized to the massive case, i.e.\ to deduce constraints on massive Feynman integrals from AdS/CFT integrability.
When mass is introduced into $\superN=4$ SYM theory via the Higgs mechanism, an extended massive version of dual (super)conformal symmetry was found to survive the symmetry breaking \cite{Alday:2009zm}. However, no such massive extension is known for the ordinary conformal symmetry and thus there is no known massive Yangian symmetry on the Coulomb branch of $\superN=4$ SYM theory. Hence, one may be tempted to believe that integrability is lost in the massive phase.
\bigskip

Recently, however, it was discovered that integrability also features in the building blocks of massive quantum field theories, i.e.\ on the very right of a potential massive version of the sequence \eqref{eq:Sequence} \cite{Loebbert:2020hxk}. Large families of ubiquitous massive Feynman integrals were found to be invariant under the generators of an extended Yangian. Again, this massive Yangian is understood as the closure of a massive dual conformal symmetry and a novel generalization of the ordinary conformal symmetry. The integrands of those families of massive Feynman integrals can be considered as the Kaluza--Klein reductions of higher dimensional massless integrands. This suggests to interpret the observed integrability properties within the AdS/CFT duality.
\bigskip

In the present paper we seek the interpretation of the massive Yangian symmetry of Feynman integrals found in \cite{Loebbert:2020hxk} as the integrability of a massive quantum field theory. Our guideline is the relation of massless integrability in $\superN=4$ SYM theory and Feynman integrals via the fishnet theory. In contrast to the massless case, we argue in the subsequent \Secref{sec:Routes} that there are (at least) two different ways to introduce interesting massive extensions of the fishnet theory. Explicitly, we distinguish the following cases:
\begin{enumerate}
\item[I.] Spontaneous symmetry breaking in the massless fishnet theory.
\item[II.] Double-scaling limit of Coulomb-branch $\superN=4$ SYM theory.
\end{enumerate}
The theories resulting from the alternative routes I and II both have their own advantages. 
The spontaneously broken fishnet (SBF) theory obtained from taking the first route was recently shown to allow for a `natural' breaking of conformal symmetry, where some of the symmetry breaking vacua survive in the quantum theory \cite{Karananas:2019fox}. Moreover, in \Secref{sec:Soft} we explicitly demonstrate that scattering amplitudes in this theory are subject to conformal soft theorems.
In \Secref{sec:Yangian} we contrast the symmetries of scattering amplitudes and Feynman integrals arising in both theories. For option II we illustrate that there is a one-to-one correspondence%
\footnote{As in the massless case, this one-to-one correspondence remains conjectural.}
 between off-shell amplitudes and those massive Feynman integrals that were found to feature a massive Yangian \cite{Loebbert:2020hxk}. Due to the close similarity to the massless bi-scalar fishnet theory, we thus refer to the theory obtained via scenario II as the massive fishnet (MF) theory. We end the paper with a summary and outlook in \Secref{sec:Conclusions}.
%%%%%%%%%%%%%%
%%%%%%%%%%%%%%%%%%%%%%%%%%%%%%%%%%%%%%%%%%%%%%%%%%%%%%%%%%%%%%%%%%%%%%%%%%%
\section{Routes to Massive Fishnets}
\label{sec:Routes}

In this section we outline two different approaches to define a massive extension of the fishnet theory. The basis for our considerations is the definition of the massless fishnet theory via a particular double-scaling limit of deformed $\superN=4$ SYM theory \cite{Gurdogan:2015csr}. We thus start with a short review of this limit, followed by the discussion of the massive cases. At the end of this section we discuss different possibilities to choose a vacuum expectation value (VEV) and to take the large-$N_\text{c}$ limit of the massive theories.

%%%%

\subsection{Review of Massless Fishnets}

The massless fishnet theories arise as the planar limits of a certain double-scaled version of gamma-deformed $\mathcal{N}=4$ SYM theory. On the level of the Lagrangian, the gamma-deformation \cite{Lunin:2005jy,Frolov:2005dj,Leigh:1995ep} can be implemented by an operator $\mathcal{P}_\gamma$ that acts bi-locally on a matrix product of some generic fields $\Phi_j$ according to%
\footnote{This formulation is reminiscent of the bi-local Yangian level-one generators.}

\begin{align}
\label{eq:biPhase}
\mathcal{P}_\gamma\brk*{\Phi_{1}\Phi_{2}\dots\Phi_{n}}
=
 \exp\brk*{\sfrac{i}{2} \gamma_A \epsilon_{ABC}\, q^B \wedge q^C}\Phi_{1}\Phi_{2}\dots \Phi_{n}.
\end{align}
Here the anti-symmetric wedge-product is defined as
\begin{align}
q^A \wedge q^B =\sum_{k=1}^n \sum_{j=1}^{k-1} \brk*{q^A_k q^B_j-q^A_j q^B_k},
\end{align}
and $q_j$ denotes the R-symmetry Cartan charge vector of the field $\Phi_j$ as given in \Appref{sec:DetailsGammaDef}.
 Under the deformation induced by the operator $\mathcal{P}_\gamma$, the terms in the Lagrangian of $\mathcal{N}=4$ SYM theory acquire different phase factors depending on the three parameters~$\gamma_{1,2,3}$. Hence, the set of parameters in the undeformed theory, i.e.\ the coupling $g$ and the number of colors~$N_\text{c}$, is enhanced by these three deformation parameters.
 
As the next step it was noticed in \cite{Gurdogan:2015csr}  that setting $\xi^2:=g^2 N_\text{c} e^{-i \gamma_3}$ and taking the double-scaling limit 
\begin{align}
g\to 0,
\qquad
\gamma_j\to i\infty,
\qquad
\xi=\text{fix},
\end{align}
the complexity of the gamma-deformed model reduces drastically. The resulting bi-scalar fishnet theory is described by the Lagrangian
\begin{equation}
\Lagr_\text{F}=N_\text{c} \tr\brk*{
-\partial_\mu \bar X \partial^\mu X 
-
\partial_\mu \bar Z \partial^\mu Z
+\xi^2 \bar X\bar Z X Z
}.
\label{eq:MasslessFishnetLagrangian}
\end{equation} 
Here $X$ and $Z$ are complex scalar fields transforming in the adjoint representation of $\grp{SU}(N_\text{c})$ and $\xi$ is the new scalar coupling.
The integrability of the fishnet theory is inherited from planar $\mathcal{N}=4$ SYM theory and materializes in its conformal and dual conformal symmetry, which combine into a Yangian \cite{Chicherin:2017cns,Chicherin:2017frs}. 

We note that at the quantum level the above massless fishnet theory is incomplete and requires the addition of further double-trace couplings \cite{Sieg:2016vap,Grabner:2017pgm}. In the present paper we aim to indentify simple massive quantum field theories by investigating the symmetries of their scattering amplitudes. At leading order in the large-$N_c$ limit the double-trace couplings do not contribute to the type of observables studied here. Therefore, we ignore them in the following.

\begin{figure}
\begin{center}
\includegraphicsbox{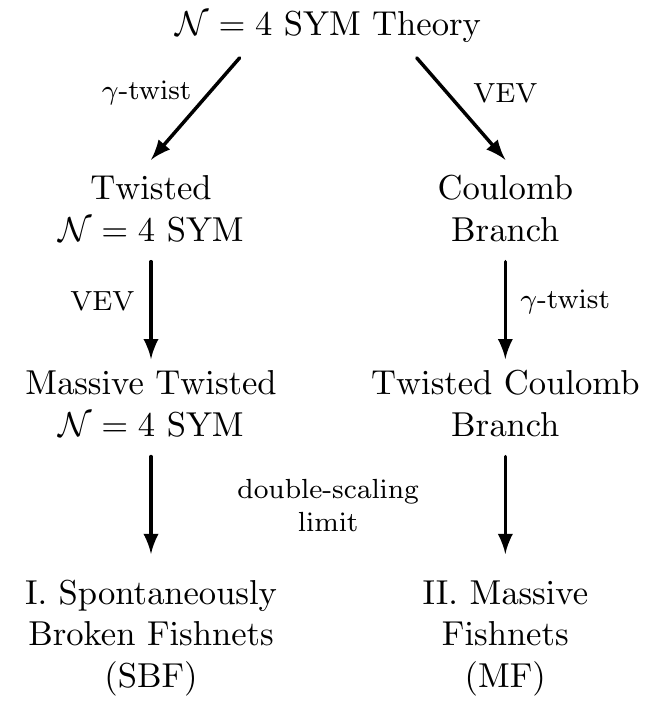}
\end{center}
\caption{Routes to massive fishnets.}
\label{Fig:RoutesToMassive}
\end{figure}

As in the case of $\mathcal{N}=4$ SYM theory, it is natural to ask what happens to the conformal and dual conformal symmetries if masses are introduced into the fishnet theory via the Higgs mechanism, cf.\ \cite{Alday:2009zm}. 
A controlled way to introduce a mass scale into a conformal theory is to spontaneously break conformal symmetry by having a scalar field acquire a non-zero vacuum expectation value (VEV). Starting from $\mathcal{N}=4$ Super-Yang-Mills theory, there are at least two distinct routes to construct a massive fishnet theory. These are illustrated in \Figref{Fig:RoutesToMassive} and discussed in detail below. These two routes differ in the order in which a scalar field acquires a vacuum expectation value and the gamma-twist is carried out.
It turns out that the double-scaling limit and the expansion around a non-zero VEV commute, thus exchanging the order of the two operations does not give rise to a third candidate theory.

%%%%%%%%%%%%%%%%%%% %%%%%%%%%%%%%%%%%%%%%%%%%%%%%%%%%%%%%%%%%%%%%%%%%%%%%%%%
\subsection{Spontaneous Symmetry Breaking in the Fishnet Theory}
\label{sec:SBFTheory}

%%%%
In this subsection we start from the double-scaled gamma-deformed theory defined by \eqref{eq:MasslessFishnetLagrangian} and let one of its scalar fields acquire a vacuum expectation value.  This situation has been investigated in the interesting letter \cite{Karananas:2019fox}. There it was shown that e.g.\ for the case of a diagonal expectation value, under certain conditions conformal symmetry survives loop corrections in perturbation theory.
Therefore, we study the case in which the field $Z$ is expanded according to%
\footnote{
Note the slight abuse of notation where the excitation on the right hand side is denoted by the same letter as the field before symmetry breaking.}
\begin{align}
\phione \rightarrow \langle \phione\rangle+\phione ,
\end{align}
where the VEV of $\phione$ takes the diagonal form
\begin{align}
{\langle \phione\rangle^a}_b &= \frac{\m_a}{\xi} \delta^a_b,
\label{eq:VEVentries}
\end{align}
with $z_a \in \mathbb{C}$.
The Lagrangian for the resulting 
spontaneously broken fishnet (SBF) theory then reads
\begin{equation}
\mathcal{L}_\text{SBF} = \mathcal{L}_\text{F} 
+ N_\text{c} \m_a \bar \m_b {\bar\phitwo^a}{}_b {\phitwo^b}_a 
+ N_\text{c} \xi \brk*{\bar \m_a
{X^a}{}_b {Z^{b}}{}_c{{\bar X}^c}{}_a
+ \m_a{\bar X^{a}}{}_b 
{\bar Z^{b}}{}_c {X^{c}}{}_a}.
\label{eq:LagrFishSpon}
\end{equation}
In \cite{Karananas:2019fox} different constraints for the entries \eqref{eq:VEVentries} of the VEV were identified in order for
conformal symmetry to stay a spontaneously broken symmetry once quantum corrections are taken into account.
Those constraints are expected to become important when extending the soft theorems of \Secref{sec:Soft}
to loop level or when considering double-trace observables.
Note that $\bar z_a$ does not necessarily have to be complex conjugate to $z_a$ which is due to the non-unitarity of the Lagrangian and the resulting equations of motion, cf.\  \cite{Karananas:2019fox}.

%%%%%%%%%%%%%%%%%%%%
\paragraph{Feynman Rules and Physical Masses.}
The propagator for the massive $\phitwo$ field is given by
\begin{equation}
\includegraphicsbox{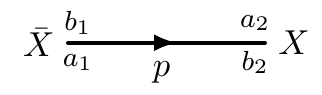} = \frac{1}{N_\text{c}}\frac{ \delta^{b_2}_{a_1}\delta^{b_1}_{a_2} 
}{p^2 +\m_{a_1} \bar \m_{b_1}}.
\label{eq:SBFpropagator}
\end{equation}
Importantly, the parameters $\m_a$ enter the propagator in the form of a product $\m_{a_1} \bar \m_{b_1}$.
Also the three-point vertices are proportional to the $\m_a$:
\begin{align}
\includegraphicsbox{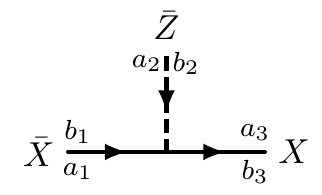} = 
 N_\text{c}\m_{a_1} \xi\, \delta^{b_3}_{a_1} \delta^{b_2}_{a_3} \delta^{b_1}_{a_2},
 \notag\\
\includegraphicsbox{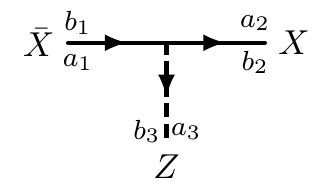}  = N_\text{c} \bar\m_{b_1} \xi\, \delta^{b_3}_{a_1} \delta^{b_2}_{a_3} \delta^{b_1}_{a_2}.\label{eq:3ptvertex}
\end{align}
For completeness, we also give the four-point vertex which takes the form
\begin{equation}
\includegraphicsbox{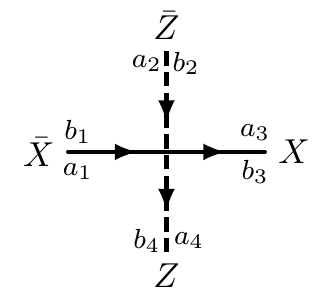} 
= N_\text{c}\xi^2\,
\delta^{b_4}_{a_1} \delta^{b_3}_{a_4} \delta^{b_2}_{a_3} \delta^{b_1}_{a_2}.
\end{equation}
It is instructive to study Feynman graphs in this theory in double-line notation.
Consider for instance the example
\begin{equation}
\includegraphicsbox{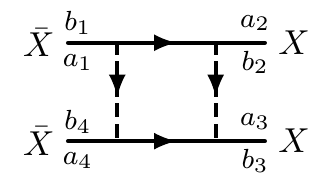}
=
\delta^{b_4}_{a_1}
\delta^{b_1}_{a_2}
\delta^{b_2}_{a_3}
\delta^{b_3}_{a_4}
\sum_{a_0=1}^{N_\text{c}}
\includegraphicsbox{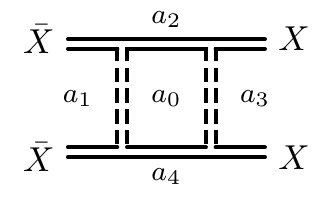}.
\end{equation}
Here it becomes clear that there is a natural association between the color index carried by a line and the region in the Feynman diagram surrounded by that line. Since the mass parameters $\m_a$ also carry a color index, they correspond to the regions and we refer to them as region masses.  The physical mass of a particle propagating between two regions can be read off from the propagator \eqref{eq:SBFpropagator} and is given by the product of the two corresponding region masses:
\begin{align}
 m_{a b}^2 = \m_a \bar \m_b.
\end{align}
The spectrum of physical masses consists of all possible products of two region masses $z_a$. In contrast, on the Coulomb branch of $\mathcal{N}=4$ SYM theory, the mass term in a propagator is given by the differences of neighboring region masses $m_a$ and $m_b$:
\begin{align}
\brk!{m_{ab}^{\mathcal{N}=4}}^2 = (m_a - m_b)^2.
\end{align}
As opposed to $z_a$ and $\bar z_a$, in the latter case the real parameters $m_a$ on their own take the role of physical masses.
It would be desirable to identify a fishnet-like theory with the latter propagator structure, since massive Feynman diagrams with such difference-mass propagators were recently found to be subject to a Yangian symmetry~\cite{Loebbert:2020hxk}. This motivates to search for an alternative massive extension of the fishnet theory. We will come back to the situation of spontaneous symmetry breaking in due course.

%%%%%%%%%%%%%%%%%%%%%%%%%%%%%
\subsection{Double-Scaling Limit of Coulomb-Branch $\mathcal{N}=4$ SYM Theory}
\label{sec:MFLimit}

In this subsection we take the second route to a massive fishnet theory that was indicated above. We show that in the simplest case, this route results in a bi-scalar theory which inherits the difference-mass propagators from the Coulomb branch of $\mathcal{N}=4$ SYM theory. As discussed in more detail in the subsequent \Secref{sec:Yangian}, off-shell amplitudes in this theory are in one-to-one correspondence with massive fishnet Feynman integrals that have been found to be Yangian-invariant, cf.\ \cite{Loebbert:2020hxk}. This motivates to refer to the resulting theory as the massive fishnet (MF) theory. At the end we will also discuss a slightly more general double-scaling limit which yields a tri-scalar theory.

Our starting point is the Euclidean Lagrangian of $\mathcal{N}=4$ SYM theory, see e.g.\ \cite{Fokken:2014soa}:
\begin{align}
\mathcal{L}_{\mathcal{N}=4}
=
N_\text{c}\tr\Big[
&-\quarter F_{\mu\nu}F^{\mu\nu}- D^\mu \phi_j^\dagger D_\mu \phi^j
+i \bar\psi_A^{\dot \alpha}D_{\dot \alpha}^\alpha \psi_\alpha^A
\nonumber\\
&
- \quarter g^2 \acomm{\phi_i^\dagger}{\phi^i}\acomm{\phi_j^\dagger}{\phi^j}
+ g^2 \phi_i^\dagger \phi_j^\dagger \phi^i\phi^j
\nonumber\\
&
+g\bar \psi_4 \comm{\phi^j}{ \bar\psi_j}
- gi\epsilon _{ijk}\psi^k\phi^i\psi^j
+g\psi_j\comm{\phi_j^\dagger}{ \psi_4}
- gi\epsilon^{ijk}\bar \psi_k\phi_i^\dagger \bar \psi_j
\Big],
\end{align}
where $i,j=1,2,3$, $A=1,2,3,4$, $D_{\dot \alpha}^\alpha=D_\mu (\tilde \sigma^\mu)_{\dot \alpha}^\alpha$ and $(\tilde \sigma^\mu)_{\dot \alpha}^\alpha=(-i \sigma_2,i\sigma_3,1,-i\sigma_1)_{\dot \alpha}^\alpha$ and 
 \begin{equation}
 D_\mu=\partial_\mu+ \frac{i g}{\sqrt{2}} \comm{A_\mu}{\cdot},
 \qquad
 F_{\mu\nu}=-\frac{i \sqrt{2}}{g}  \comm{D_\mu}{D_\nu}.
 \end{equation}
Note that the fields in the above Lagrangian have been rescaled by a factor of $\sqrt{N}_\text{c}$ and we have $g^2=g_\text{YM}^2N_\text{c}$.%
\footnote{Often one defines $\lambda=g_\text{YM}^2N_\text{c}$ and $\tilde g^2=g_\text{YM}^2N_\text{c}/(4\pi^2)$, which may result in factors of $4\pi$ when the results are expressed in terms of $\tilde g$.}

%%%%%%
\paragraph{Introducing a VEV.}

We give a \emph{real} VEV to the complex scalar field $\phi_3$, i.e.\ we replace
\begin{equation}
\phi_3 \to 
 \langle\phi_3\rangle +\phi_3,
\end{equation}
with the real VEV matrix $V:= \langle\phi_3\rangle=\langle\phi_3\rangle^*$.%
\footnote{
This reality condition implies that only one of the six real scalars of $\superN=4$ SYM theory acquires a VEV, cf.\ \cite{Alday:2009zm}. Here we denote the VEV entries by $m_a$ since they correspond to physical masses of the theory. In contrast, in \Secref{sec:SBFTheory} only the products of the VEV entries $z_a$ yield physical masses.}
Moreover, we choose the VEV to be diagonal:
\begin{equation}
V= g^{-1} \Lambda,
\qquad
\Lambda=\text{diag}(m_1,\dots,m_{N_\text{c}}).
\end{equation}
Here some of the $m_a$ can and will eventually be set to zero, which is convenient to adjust the types of Feynman diagrams with massive and massless propagators that contribute to observables in the planar limit.
Due to their different roles we denote the first two complex matrix fields by
\begin{equation}
X=\phi_1,
\quad
 \bar X=\phi_1^\dagger,
\qquad\qquad
Z=\phi_2, 
\quad
\bar Z=\phi_2^\dagger,
\end{equation}
and we introduce the Hermitian matrices
\begin{equation}
Y=\sfrac{1}{\sqrt{2}}(\phi_3+\phi_3^\dagger),
\qquad
W=-\sfrac{1}{\sqrt{2}} i(\phi_3-\phi_3^\dagger).
\end{equation}
Then we can write the Coulomb-branch Lagrangian of $\superN=4$ SYM theory as
\begin{align}
\mathcal{L}_\text{Coul}
=
\mathcal{L}_{\mathcal{N}=4}+
\mathcal{L}_\text{VEV},
\label{eq:LCoul}
\end{align}
where we have defined
\begin{align}
\mathcal{L}_\text{VEV}
&=
 N_\text{c}\tr\Big[
\half g^2\comm{V}{W}^2
+g^2\comm{V}{\phi_\primed{i}}\comm{V}{\phi_\primed{j}^\dagger}
+ \half g^2\comm{V }{A^\mu}^2
+ig\comm{V }{A^\mu}\partial_\mu Y
\nonumber\\
&
+\sqrt{2}g^2 \brk*{
\sfrac{1}{2}\comm{V}{W}\comm{Y}{W}
+\half \comm{V }{A^\mu}\comm{Y}{A_\mu}
- \half \acomm{V}{Y}\acomm{\phi_{j'}}{\phi_{j'}^\dagger}
+V\phi_\primed{j}Y\phi_\primed{j}^\dagger
+V \phi_\primed{j}^\dagger Y\phi_\primed{j}
}
\nonumber\\
&
+g\bar \psi_4 \comm{V}{\bar\psi_3}
-gi\epsilon _{3\primed{j}\primed{k}}\psi^\primed{k}V\psi^\primed{j}
+g\psi_3 \comm{V}{\psi_4}
-gi\epsilon^{3\primed{j}\primed{k}}\bar \psi_\primed{k}V \bar \psi_\primed{j}
\Big].
\label{eq:LCoulVEV}
\end{align}
Here the indices $\primed{i},\primed{j}$ are summed from 1 to 2.
Next we remove the mixing between the scalar $Y$ and the vector $A^\mu$ by adding the $R_{\xi=\alpha}$ gauge fixing term, cf.\ \cite{Alday:2009zm}%
\footnote{This is particularly important for the double-scaling limit considered below, where $Y$ only decouples if this mixing term is removed.}
\begin{equation}
\Lagr_\text{gauge}
=
-\half N_\text{c} \tr G^2,
\qquad
G=\frac{1}{\sqrt{\alpha}}
\brk*{
\partial_\mu A^\mu+ig \alpha \comm{V}{Y}
},
\end{equation}
and we add the ghost term
\begin{equation}
\Lagr_\text{ghost}
=
\tr\brk*{
\bar c\partial^\mu D_\mu c
-g^2 \alpha \, \bar c\comm{V}{\comm{V+Y}{c}}
}.
\end{equation}
This yields the final VEV Lagrangian
\begin{align}
\mathcal{L}_\text{VEV}^{\text{gg}}
&=
 N_\text{c}\tr\Big[
\half g^2\comm{V}{W}^2
+g^2\comm{V}{\phi_\primed{i}}\comm{V}{\phi_\primed{j}^\dagger}
+ \half g^2\comm{V }{A^\mu}^2
+ \half g^2 \alpha \comm{V}{Y}^2
\nonumber\\
&
+\sqrt{2}g^2 \brk*{
\sfrac{1}{2}\comm{V}{W}\comm{Y}{W}
+\half \comm{V }{A^\mu}\comm{Y}{A_\mu}
- \half \acomm{V}{Y}\acomm{\phi_{j'}}{\phi_{j'}^\dagger}
+V\phi_\primed{j}Y\phi_\primed{j}^\dagger
+V \phi_\primed{j}^\dagger Y\phi_\primed{j}
}
\nonumber\\
&
+g\bar \psi_4 \comm{V}{\bar\psi_3}
-gi\epsilon _{3\primed{j}\primed{k}}\psi^\primed{k}V\psi^\primed{j}
+g\psi_3 \comm{V}{\psi_4}
-gi\epsilon^{3\primed{j}\primed{k}}\bar \psi_\primed{k}V \bar \psi_\primed{j}
\nonumber\\
&
-\frac{1}{2\alpha} (\partial_\mu A^\mu)^2
+
\bar c\partial^\mu D_\mu c
-g^2 \alpha \, \bar c\comm{V}{\comm{V+Y}{c}
}
\Big].
\label{eq:LCoulVEV2}
\end{align}
Note that the scalar field $Y$ has acquired an $\alpha$-dependent mass from the above gauge fixing contribution.
In the following we will set $\alpha=1$. 
Before moving on to deforming this Lagrangian, we note that in components it is easier to identify the mass terms of difference form. For $V^a{}_b=g^{-1}\, m_a \delta^{a}_b$ we have for instance
\begin{equation}
  \half g^2\tr\comm{V}{W}^2
=
-\half (m_a-m_b)^2 W^a{}_b W^b{}_a.
\end{equation}

%%%%%%%%
\paragraph{Gamma-Deformation.}
We would like to mimic the derivation of the massless fishnet theory, i.e.\ to follow the sequence
\begin{equation}
\mathcal{N}=4 \text{ SYM} 
\quad\to\quad
\gamma\text{-deformation}
\quad\to\quad
\text{fishnet theory}.
\end{equation}
Hence, the next step would be to introduce a gamma-deformation by acting on the Lagrangian with the bilocal phase operator \eqref{eq:biPhase}:
\begin{align}
\mathcal{P}_\gamma\brk*{\Phi_{1}\Phi_{2}\dots\Phi_{n}}
=
 \exp\brk*{\sfrac{i}{2} \gamma_A \epsilon_{ABC}\, q^B \wedge q^C}\Phi_{1}\Phi_{2}\dots \Phi_{n}.
\end{align}
However, the action of this bi-local phase operator on a trace of fields is only well defined if this trace is invariant under R-symmetry. The traces contributing to the Lagrangian of massless $\mathcal{N}=4$ SYM theory are R-symmetry singlets, such that the action of the operator $\mathcal{P}_\gamma$ is unique, cf.\ e.g.\ \cite{Sogaard:2011aa}. In contrast, this is generically not the case for the traces contributing to the above Coulomb-branch Lagrangian. We thus refine the procedure as follows. First we introduce the partition operator $Q_{a_j}$ that maps a trace of fields to
a linear combination of non-cyclic monomials%
\footnote{This is natural from the spin chain point of view, where field traces are considered as spin chain states (monomials) with an extra cyclicity condition. See \cite{Beisert:2018ijg} for similar notation. Introducing the parameters $a_k$ may seem ad hoc at first sight. However, below we will see that these parameters drop out after taking the double-scaling limit that defines the massive fishnet theory. In principle, one should include a grading for fermionic fields into the definition of $Q_{a_j}$. Here we refrain from making this explicit since all fermions decouple in the limits we are interested in.}
\begin{equation}
\label{eq:Qmap}
Q_{a_j}: \tr(\Phi_1\Phi_2\dots \Phi_n) \mapsto
a_1 \Phi_1\Phi_2\dots \Phi_n
+a_2 \Phi_2\Phi_3\dots \Phi_1
+\dots
+a_n \Phi_n\Phi_1\dots \Phi_{n-1},
\end{equation}
with new parameters $a_k$ which obey $\sum_{k=1}^n a_k=1$. Here the only cyclically invariant choice that makes the above map well defined is $a_k=1/n$ for $k=1,\dots,n$, but it is useful to keep the parameters explicit for the moment to see that they drop out in certain limits.
Acting on the monomials on the right hand side, we then define the linear map $Q^{-1}$ such that
\begin{equation}
Q^{-1}: \Phi_1\Phi_2\dots \Phi_n \mapsto \tr(\Phi_1\Phi_2\dots \Phi_n).
\end{equation}
We thus have
\begin{equation}
Q^{-1}Q_{a_j} \tr(\Phi_1\Phi_2\dots \Phi_n)=\tr(\Phi_1\Phi_2\dots \Phi_n).
\end{equation}
The phase operator $P_\gamma$ is well defined on the above monomials and we define the twisted Coulomb-branch Lagrangian as
\begin{equation}
\mathcal{L}_\text{Coul}^{\gamma,a_j}
=
Q^{-1}\mathcal{P}_\gamma Q_{a_j}\mathcal{L}_\text{Coul}.
\end{equation}
When acting on the Lagrangian of massless $\mathcal{N}=4$ SYM theory, the insertion of the operator $Q$ makes no difference and we have
\begin{equation}
\mathcal{L}_{\mathcal{N}=4}^{\gamma}=Q^{-1}\mathcal{P}_\gamma Q_{a_j}\mathcal{L}_{\mathcal{N}=4}=\mathcal{P}_\gamma\mathcal{L}_{\mathcal{N}=4}.
\end{equation} 
Hence, by construction this modification of the gamma-deformation only affects the 
massive VEV part \eqref{eq:LCoulVEV} of \eqref{eq:LCoul}:
\begin{equation}
\mathcal{L}_\text{Coul}^{\gamma,a_j}
=
\mathcal{L}_{\mathcal{N}=4}^\gamma + Q^{-1}\mathcal{P}_\gamma Q_{a_j} \mathcal{L}_\text{VEV}.
\end{equation}
Explicitly, we have in the massless sector (see \cite{Fokken:2013aea, Caetano:2016ydc})
\begin{align}
\mathcal{L}_{\mathcal{N}=4}^\gamma
=
N_\text{c}\tr\Big[
&-\quarter F_{\mu\nu}F^{\mu\nu}- D^\mu \phi_j^\dagger D_\mu \phi^j
+i \bar\psi_A^{\dot \alpha}D_{\dot \alpha}^\alpha \psi_\alpha^A
\nonumber\\
&
- \quarter g^2 \acomm{\phi_i^\dagger}{\phi^i}\acomm{\phi_j^\dagger}{\phi^j}
+ g^2 e^{-i \eps^{ijk}\gamma_k}\phi_i^\dagger \phi_j^\dagger \phi^i\phi^j
\nonumber\\
&
-ge^{-\ihalf \gamma_j^-}\bar \psi_j \phi^j \bar \psi_4+ge^{+\ihalf \gamma_j^-}\bar \psi_4 \phi^j \bar\psi_j
- ig\epsilon _{ijk} e^{\ihalf \epsilon_{jkm}\gamma_m^+}\psi^k\phi^i\psi^j
\nonumber\\
&-
ge^{+\ihalf \gamma_j^-}\psi_4 \phi_j^\dagger \psi_j
+ge^{-\ihalf \gamma_j^-}\psi_j\phi_j^\dagger \psi_4
- ig\epsilon^{ijk}e^{\ihalf \epsilon_{jkm}\gamma_m^+}\bar \psi_k\phi_i^\dagger \bar \psi_j
\Big],
\end{align}
where one defines
\begin{equation}
\gamma_1^\pm=-\frac{\gamma_3\pm\gamma_2}{2},
\qquad
\gamma_2^\pm=-\frac{\gamma_1\pm\gamma_3}{2},
\qquad
\gamma_3^\pm=-\frac{\gamma_2\pm \gamma_1}{2}.
\end{equation}
We note that at the quantum level completeness of the theory defined by the massless Lagrangian $\mathcal{L}_{\mathcal{N}=4}^\gamma$ requires additional double-trace counter terms, cf.\ \cite{Dymarsky:2005uh,Pomoni:2008de,Fokken:2013aea,Fokken:2014soa}. Here our focus lies on making the first step in identifying an integrable massive quantum field theory. We thus ignore these extra contributions, and leave an exploration of the completeness of the considered massive theories for future work.

Similarly to the massless sector, we may evaluate the gamma-deformation of the above VEV contribution
\begin{equation}
\mathcal{L}_\text{VEV}^\gamma=Q^{-1}\mathcal{P}_\gamma Q_{a_j} \mathcal{L}_\text{VEV}.
\label{eq:LVEVgamma}
\end{equation}
While it is straightforward to obtain $\mathcal{L}_\text{VEV}^\gamma$ using computer algebra, the explicit expression for this intermediate result is lengthy and not very illuminating.
Keeping the parameters $a_j$ in \eqref{eq:Qmap} explicit, in the first of the below limits, these parameters will drop out of the Lagrangian. We will then also consider a limit in which the respective parameters remain in the resulting theory.
\paragraph{Massive Fishnet Theory.}

Next we want to take a double-scaling limit in the spirit of the massless discussion of~\cite{Gurdogan:2015csr}. We define the new coupling
\begin{equation}
\xi^2=g^2  e^{-i \gamma_3},
\end{equation}
and take the limit%
\footnote{For $\gamma_{1,2}$ finite this implies $\xi_{1,2}\to 0$ with $\xi_j^2=g^2 e^{-i \gamma_j}$}
\begin{align}
g\to 0,
\qquad
\gamma_3\to i\infty,
\qquad
\xi=\text{fix}.
\label{eq:DoubleLimit1}
\end{align}
As shown in \cite{Gurdogan:2015csr}, in the massless sector this yields the bi-scalar fishnet Lagrangian
(modulo kinetic terms for fields that decouple from the theory) 
\begin{equation}
\Lagr_\text{F}
=
N_\text{c}\tr\brk*{
-\partial_\mu \bar X \partial^\mu X 
-
\partial_\mu \bar Z \partial^\mu Z
+ \xi^2 \bar X\bar Z X Z
}.
\label{eq:DoubleLimitIMasslessLagr}
\end{equation} 
The full Lagrangian to which we refer as the \emph{massive fishnet theory} Lagrangian takes the form
\begin{equation}
\Lagr_\text{MF}=\Lagr_\text{F}+\mathcal{L}_\text{M},
\label{eq:FullLagrBiI}
\end{equation}
where for the VEV part of the gamma-deformed Lagrangian \eqref{eq:LVEVgamma} the above limit yields
\begin{align}
\mathcal{L}_\text{M} 
&= 
 N_\text{c}\tr\brk*{ \comm{\Lambda}{X}\comm{\Lambda}{\bar X}+ \comm{\Lambda}{Z}\comm{\Lambda}{\bar Z}}
 \nonumber\\
 &=-  N_\text{c}(m_a - m_b)^2 {X^a}_b{\bar X^b}_a
- N_\text{c}(m_a - m_b)^2 {Z^a}_b{\bar Z^b}_a.
\end{align}
Here in going to the second line we used the explicit form of the diagonal VEV matrix $\Lambda^a{}_b=m_b \delta_b^a$.
We remark that taking the planar limit of this theory corresponds to taking $N_\text{c}\to \infty$ while keeping $\xi$ and the masses $m_a$ fix.
As opposed to the SBF Lagrangian \eqref{eq:LagrFishSpon}, the above MF Lagrangian \eqref{eq:FullLagrBiI} features mass terms for the two scalars but no three-point vertices. 

Note that if in addition we take $\gamma_{1,2}\to i\infty$ with $\xi_{j}=\xi^2=g^2  e^{-i \gamma_j}$ fix (now also for $j=1,2$ in \eqref{eq:DoubleLimit1}), some contributions to the VEV Lagrangian scale as $(\text{finite})/g$ and diverge in the limit $g\to 0$. This differs from the massless case where the respective limit yields a finite theory. In the massive case, however, there are alternative ways to the above limit which keep these terms finite. We illustrate this below for a tri-scalar example.

For completeness, we also give the Feynman rules for the massive fishnet theory. 
The propagator, connecting a conjugate field $\bar X$ or $\bar Z$ on the left with a field $X$ or $Z$ on the right, respectively, reads
\begin{equation}
\includegraphicsbox{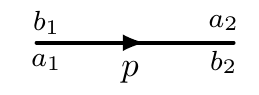} = \frac{1}{N_\text{c}}\frac{ \delta^{b_2}_{a_1}\delta^{b_1}_{a_2} 
}{p^2 +(m_{a_1}- m_{b_1})^2}.
\label{eq:MFpropagator}
\end{equation}
Moreover, the four-point vertex takes the form
\begin{equation}
\includegraphicsbox{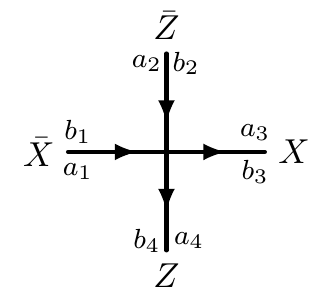} 
= N_\text{c}\xi^2\,
\delta^{b_4}_{a_1} \delta^{b_3}_{a_4} \delta^{b_2}_{a_3} \delta^{b_1}_{a_2}.
\end{equation}
If we consider the VEV scenario described in \Secref{sec:PlanarAndVEV}, only some of the masses $m_a$ are non-vanishing such that also massless propagators contribute to the resulting observables.

%%%%%%%%%%%%%%%%
\paragraph{Tri-Scalar Double-Scaling Limit.}

To take an alternative limit we define three couplings $\xi_j$ differently and according to
\begin{equation}
\xi_1^2=g e^{-i \gamma_1},
\qquad
\xi_2^2=g e^{-i \gamma_2},
\qquad
\xi_3^2=g^2  e^{-i \gamma_3},
\end{equation}
and then take the limit
\begin{align}
g\to 0,
\qquad
\gamma_j\to i\infty,
\qquad
\xi_j=\text{fix}.
\end{align}
For the resulting Lagrangian $\Lagr^\text{Tri}=\Lagr_{\text{F}}^\text{Tri}+\mathcal{L}_\text{M}^\text{Tri}$,
the massless sector takes the form 
\begin{align}
\Lagr_{\text{F}}^\text{Tri}
=
& N_\text{c}\tr\brk*{
-\partial_\mu \bar X \partial^\mu X 
-
\partial_\mu \bar Z \partial^\mu Z
-
\half \partial_\mu Y \partial^\mu Y
+ \xi_3^2 \bar X\bar Z X Z
},
\end{align}
while the massive contribution becomes
\begin{align}
\mathcal{L}_\text{M}^\text{Tri} = 
 N_\text{c}\tr\Big[
&\comm{\Lambda}{X}\comm{\Lambda}{\bar X}
+\comm{\Lambda}{Z}\comm{\Lambda}{\bar Z}
+ \half \comm{\Lambda}{Y}^2
\nonumber\\
&+
\frac{\xi_1^2}{\sqrt{2}}\brk*{
c_1 \Lambda Y\bar Z Z
+c_2\Lambda\bar ZZY
+c_3\Lambda ZY\bar Z
}
\nonumber\\
&+
\frac{\xi_2^2}{\sqrt{2}}\brk*{
c_1 \Lambda YX\bar X
+c_2 \Lambda X\bar X Y
+c_3\Lambda \bar X Y X
}\Big].
\end{align}
\begin{table}
\begin{center}
\begin{tabular}{|l|c|c|c|}\hline
Theory & 4pt Vertex & 3pt Vertex & Propagators
\\\hline\hline
Spontaneously Broken Fishnets (SBF) & Yes & Yes & Product
\\\hline
Massive Fishnets (MF) & Yes & No & Difference
\\\hline
Tri-Scalar Theory& Yes & Yes & Difference
\\\hline
\end{tabular}
\end{center}
\caption{Comparison of the different massive extensions of the fishnet theory discussed here.}%
\label{tab:Comparison}%
\end{table}%
Here the first line contains the difference-mass terms and the parameters $c_k$ correspond to rescaled combinations of the parameters $a_j$ introduced via the map \eqref{eq:Qmap}. For a trace of length $4$ the cyclic choice is $a_j=1/4$ which implies 
\begin{equation}
c_k=\sfrac{1}{4}(-1,-1,+4)_k,
\qquad k=1,2,3.
\end{equation}
We remark that taking the same limit in the purely massless case with $\Lambda=0$, the scalar $Y$ decouples from the theory whereas in the massive case considered here, it interacts with the scalars $X$ and $Z$ via three-point vertices. 

We note that the ladder diagrams of Wick and Cutkosky, which have been known to feature a dual conformal symmetry for a long time \cite{Wick:1954eu,Cutkosky:1954ru}, are naturally embedded into this theory which includes a massless and two massive scalars. These ladder graphs have massive stringers and massless rungs taking the form%
\footnote{For a judicious choice of VEV configuration, as explained in the subsequent section \Secref{sec:PlanarAndVEV}, the Wick--Cutkosky ladders are the only contributions to the scattering amplitude $A(\bar X(m_1),\bar Z(m_2),Z(m_2), X(m_1))$ in the tri-scalar theory.
}
\begin{equation}
\includegraphicsbox{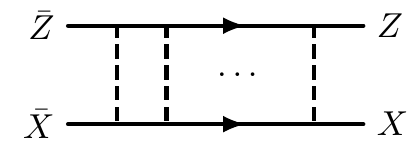}
\end{equation}
In conclusion, the massive extensions of the fishnet theory considered in this and in the previous subsection compare as in \Tabref{tab:Comparison}.

%%%%%%%%%%

\subsection{Planar Limit and VEV Scenarios}
\label{sec:PlanarAndVEV}

In the massive phases introduced above, taking the large-$N_\text{c}$ limit necessitates some care. In fact, while we will keep referring to this limit as the `planar limit', the surviving diagrams at leading order of a 
${1}/{N_\text{c}}$ expansion are not necessarily all (geometrically) planar diagrams of the theory. 
We consider a diagonal VEV matrix as discussed above:
\begin{equation}
V= \text{diag}(v_1,\dots,v_{N_\text{c}}).
\end{equation}
Now there are different possibilities for the scaling of the VEV entries $v_j$ with $N_\text{c}$. A priori, for every $v_j$, we can distinguish three different behaviors:
\begin{align}
\lim_{N_\text{c}\rightarrow \infty} v_i(N_\text{c}) \rightarrow \left\{\begin{array}{c} 0 \\ \text{finite} \\ \infty \end{array}\right. .
\end{align}
The degrees of freedom related to diverging VEV entries will be non-dynamical and decouple from the theory. Hence, we are effectively limited to the first two scaling options. In principle, we can then consider arbitrary configurations of vanishing and non-vanishing region masses. 
However, here we are interested in configurations that inherit as much symmetry from the original massless planar theory as possible. This singles out a particular subset of configurations, which have finitely many non-vanishing and infinitely many vanishing VEV entries in the planar limit:
\begin{equation}
V=\text{diag}(v_1,\dots,v_n,0,0\dots,0).
\label{eq:DistinguishedVEV}
\end{equation}

%%%
%%%%%%%

\paragraph{SBF Theory.}

\begin{figure}
\begin{center}
\includegraphicsbox{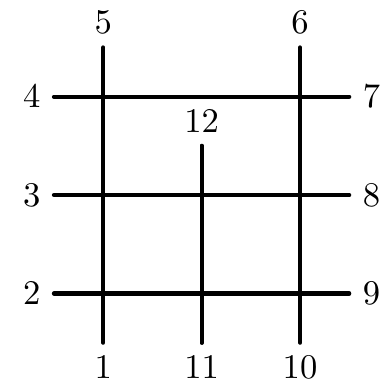}\qquad $\longrightarrow$ \qquad\includegraphicsbox{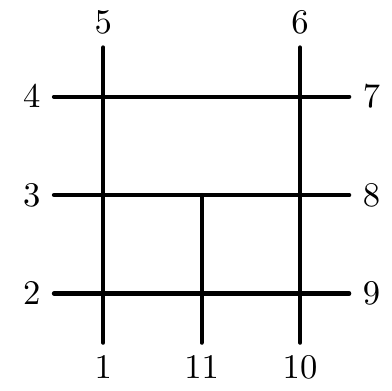}
\end{center}
\caption{Replacing one external leg of a non-planar 12-point diagram by a VEV turns it into a planar 11-point diagram.}
\label{Fig:PseudoPlanar}
\end{figure}

 To illustrate why this configuration is distinguished in the case of spontaneous symmetry breaking, consider the example of the 12-point diagram in \Figref{Fig:PseudoPlanar} which contributes to the non-planar sector of the massless theory. 
While the left-hand diagram is non-planar, replacing the external leg 12 by a VEV turns it into a (geometrically) planar diagram, which for generic VEV configurations might contribute to observables in the large-$N_\text{c}$ limit of the massive theory (spoiling e.g.\ the dual conformal symmetry). We refer to this class of diagrams as pseudo-planar, because they arise from Higgsing non-planar diagrams. To argue how the VEV configurations, which we singled out above, remove all pseudo-planar diagrams, let us first establish a canonical normalization for scattering amplitudes in the planar limit. We denote the canonically normalized $n$-point amplitude for the scattering of $n$ particles of type~$\Phi_a$ by\footnote{While this definition in terms of $S$-matrix elements of one-particle-states is valid in the massive phase, in the massless phase the right-hand-side is to be understood as a formal object constructed via the LSZ-reduction, cf. \cite{Gillioz:2020mdd}.}
\begin{align}
\langle \Phi_1 \Phi_2...\Phi_n\rangle =  N_\text{c}^{-1} {}_\text{out}\langle 0|S|\Phi_1,...\Phi_n\rangle_\text{in},
\label{eq:CanNorm}
\end{align} 
as well as its planar limit by
\begin{align}
\langle \Phi_1 \Phi_2 ... \Phi_n\rangle_\text{planar} = \lim_{\substack{N_\text{c}\rightarrow \infty\\ \xi^2 = \text{const.}}} \langle \Phi_1 \Phi_2 ... \Phi_n\rangle.
\end{align} 
Here we consider the scattering of individual component fields, i.e.\ the $\Phi_a$ carry fundamental and anti-fundamental $\grp{SU}(N_\text{c})$ indices. It is easy to see that in the massless phase, this reproduces the usual definition of the planar limit where only geometrically planar diagrams, i.e.\ diagrams which can be drawn on a plane without self-intersections, contribute to planar amplitudes. 

Now note that there are no pseudo-planar diagrams with all internal regions massless in the SBF theory. This is due to the fact that the three-point vertex (\ref{eq:3ptvertex}) is proportional to the adjacent region mass and vanishes if that region is massless. If there are only finitely many non-vanishing region masses, the planar limit will be dominated by diagrams with vanishing region masses and therefore only proper planar diagrams will contribute. In particular, the canonical normalization \eqref{eq:CanNorm} ensures that observables, whose leading contribution in a ${1}/{N_\text{c}}$ expansion consists of pseudo-planar diagrams only, vanish in the planar limit.

%%%%%%%

\paragraph{Massive Fishnet Theory.}

Also for the massive fishnet theory introduced in \Secref{sec:MFLimit} the above VEV scenario is preferred: finitely many region masses take non-vanishing values, while infinitely many vanish, at least in the planar limit.  In the MF theory this is advantageous since for this choice only diagrams with massive propagators on the boundary of the Feynman graph contribute to planar scattering amplitudes. These are precisely the Feynman graphs that feature the massive Yangian symmetry \cite{Loebbert:2020hxk}. In the following we will thus restrict to this VEV scenario and we will usually assume the first $n$ VEV entries to be non-vanishing, cf.\ \eqref{eq:DistinguishedVEV}. 

\section{Integrability alias Yangian Symmetry}
\label{sec:Yangian}

It is well known that planar $\mathcal{N}=4$ SYM theory owes many of its interesting properties to the interplay of its conformal and dual conformal symmetry, which combine to form an infinite dimensional symmetry of Yangian type. 
Here we aim at studying the symmetry properties of observables that are well defined in massless and massive theories, since we want to understand how integrability carries over from the massless to the massive situation.
The objects we will consider are planar color-ordered amplitudes, i.e.\  the leading coefficients multiplying the different tensor structures in the expansion\footnote{The form of  this expansion follows from the structure of the double-line Feynman rules.}
\begin{align}
\langle (\Phi_1)^{a_1}_{b_1} (\Phi_2)^{a_2}_{b_2} ... (\Phi_n)^{a_n}_{b_n}\rangle = \sum_{\sigma \in S_n/Z_n} \delta^{a_{\sigma(2)}}_{b_{\sigma(1)}}\delta^{a_{\sigma(3)}}_{b_{\sigma(2)}} ... \delta^{a_{\sigma(1)}}_{b_{\sigma(n)}} A_n(\Phi_{\sigma(1)}, ... \Phi_{\sigma(n)}),
\label{eq:colorOrderExp}
\end{align}
where $\vec{\Phi}\in \{X,\bar X, Z, \bar Z\}^n$ and we suppressed the momentum dependence of the external fields. 
In massless theories (or theories where all fields have the same mass) a customary way of projecting ordinary amplitudes on one particular color-ordered amplitude is by taking a trace over the external color indices. In fact, from the above expansion it is easy to see that 
\begin{align}
A_n(\Phi_1,...,\Phi_n) = \frac{1}{N_c^n} \left(\langle \tr(\Phi_1 \Phi_2 ... \Phi_n)\rangle+\mathcal{O}(\sfrac{1}{N_c})\right).
\end{align}
For the massive theories introduced above, this identification is not possible since the $A_n$ in \eqref{eq:colorOrderExp} still depend on the color indices through the region masses. That is, taking the trace leads to a sum over external particles with different masses, whereas every term in \eqref{eq:colorOrderExp} has well defined external masses. 
In the present paper we will therefore not consider single-trace amplitudes but rather read off the color-ordered amplitudes from the expansion \eqref{eq:colorOrderExp}.

%%%%%%%%%
\subsection{Review of Massless Fishnets}

Before we discuss how Yangian symmetry works in the massive theories described above, let us briefly review the way it is realized in the massless fishnet theory. 
In \cite{Chicherin:2017cns} it was shown that in the planar limit the color-ordered amplitudes defined by \eqref{eq:colorOrderExp} feature a (dual) conformal Yangian symmetry. Furthermore, since only a single Feynman diagram contributes to every amplitude, this symmetry extends to the diagrammatic level and can be used to bootstrap single Feynman diagrams without reference to a particular quantum field theory \cite{Loebbert:2019vcj}.

To exhibit the Yangian symmetry, the amplitudes are written in terms of dual position variables, the so-called region momenta $x_j$. These are related to the original momenta of the momentum space amplitude via
\begin{align}
p_j = x_{j}-x_{j+1}.\label{eq:momentummap}
\end{align} 
In these variables, the Yangian symmetry of amplitudes corresponds to invariance under the action of a set of local and bilocal generators given by%
\footnote{Note that in addition the definition of the Yangian algebra requires the so-called Serre relations for nested commutators of these generators, cf.\ \cite{Bernard:1992ya,MacKay:2004tc,Torrielli:2011gg,Loebbert:2016cdm}.}
\begin{align}
\gen{J}^A = \sum_{k=1}^n \gen{J}_k^A, && \widehat{\gen{J}}^A = \frac{1}{2} f^A{}_{BC} \sum_{j<k=1}^n \gen{J}_k^C \gen{J}_j^B + \sum_{j=1}^n s_j \gen{J}_j^A,
\label{eq:DCGenerators}
\end{align}
where $\gen{J}^A \in \{\gen{P}^\mu, \gen{L}^{\mu\nu}, \gen{D}, \gen{K}^\mu\}$ denotes the conformal Lie algebra generators with the local currents 
\begin{align}
\gen{P}_i^\mu &= -i \frac{\partial}{\partial x_{i\mu}}, 
& \gen{K}_{i}^\mu &= -2ix_i^\mu\left(x_i^\nu  \frac{\partial}{\partial x_i^\nu}  + \Delta_i\right) +i x_i^2 \frac{\partial}{\partial x_{i\mu}},\notag\\
\gen{L}_i^{\mu\nu} &=i\left(x_i^\mu \frac{\partial}{\partial x_{i\nu}} - x_i^\nu \frac{\partial}{\partial x_{i\mu}}\right), &
\gen{D}_i &= -i \left(x_i^\mu \frac{\partial}{x_i^\mu}  + \Delta_i\right).
\label{eq:DCCurrents}
\end{align}
In \eqref{eq:DCGenerators}, the $f^A{}_{BC}$ are the inverse structure constants of the conformal Lie algebra formed by the local currents. The $s_j$ represent the so-called evaluation parameters and their rational values depend on the specific amplitude under consideration. The generators $\gen{J}^A$ and $\widehat{\gen{J}}^A$ are called level-zero and level-one dual conformal generators, respectively. For example, the level-one momentum generator of the conformal algebra explicitly reads
\begin{align}
\widehat{\gen{P}}^\mu &= \frac{i}{2} \sum_{j,k=1}^n \text{sign}(j-k)\brk*{\gen{D}_j \gen{P}_k^\mu  + \gen{L}_j^{\mu\nu} \gen{P}_{k\nu}}  + \sum_{j=1}^n s_j \gen{P}_j^\mu.
\end{align}
The other level-one generators are detailed in \Appref{app:Level1Generators}. To summarize, the Yangian symmetry implies a set of differential equations that color-ordered amplitudes satisfy, namely
\begin{align}
\gen{J}^A A_n = 0, \qquad \widehat{\gen{J}}^A A_n = 0.
\end{align}
Note that for on-shell legs conformal symmetries may be broken in a subtle way that requires further investigation beyond the scope of the present paper, cf.\ \cite{Bargheer:2009qu,Sever:2009aa,Bargheer:2011mm,Bargheer:2012cp,Bianchi:2012cq,Chicherin:2017bxc}. Here we will focus on the properties of off-shell amplitudes as the criterion to single out a theory with distinguished symmetry properties.

 In \cite{Alday:2009zm} it was found that even though its conformal symmetry is spontaneously broken, $\mathcal{N}=4$ SYM theory on the Coulomb branch still features a massive representation of dual conformal symmetry. This representation can be understood as the ordinary massless representation of conformal symmetry in five spacetime dimensions with the fifth component of the vector $x^\mu_j$ playing the role of the mass $m_j$. Motivated by this result, in the following we study the dual conformal properties of two of the massive models we described above: the spontaneously broken fishnet theory and the massive fishnet theory. 
 %%%%%%%%%%%%%%
 
\subsection{Spontaneous Symmetry Breaking}
\label{sec:SymsSBF}

We find that the level-zero dual conformal symmetry of the massless theory carries over directly to the theory with spontaneously broken conformal symmetry (the SBF theory) by simply modifying the scaling and special conformal generator to read
\begin{align}
\gen{D}_i &= -i \brk*{x_i^\nu \frac{\partial}{\partial x_i^\nu} + \m_i \frac{\partial}{\partial \m_i}+ \bar{\m}_i \frac{\partial}{\partial \bar{\m}_i} + \Delta_i}, \notag\\
 \gen{K}_i^\mu &= -2 i x_i^\mu \brk*{x_i^\nu \frac{\partial}{\partial x_i^\nu} + \m_i \frac{\partial}{\partial \m_i} +\bar{\m}_i \frac{\partial}{\partial \bar{\m}_i}+ \Delta_i} +i x_i^2 \frac{\partial}{\partial x_{i\mu}}.
\end{align}
Note that these generators differ from those in the five-dimensional formulation of Coulomb-branch $\mathcal{N}=4$ SYM in \cite{Alday:2009zm}.%
\footnote{The form of the dilatation operator agrees with the six-dimensional formulation of \cite{Dennen:2010dh}, where $z$ and $\bar z$ map to the fifth and sixth components of an extended spacetime vector.} 
Still, these generators satisfy the same algebra, thereby forming a representation of dual conformal symmetry.

%%%%%%

Of course, Poincar\'e symmetry, i.e.\ invariance under the action of $\gen{P}^\mu$ and $\gen{L}^{\mu\nu}$ as given in \eqref{eq:DCCurrents} is manifest. The key to conformal symmetry is the covariant behavior of the product mass propagator, i.e.\
\begin{align}
\left(\gen{D}_1 + \gen{D}_2\right) \frac{1}{x_{12}^2+ \m_1 \bar{\m}_2} &= \frac{i(2-\Delta_1 - \Delta_2)}{x_{12}^2 + \m_1\bar{\m}_2},
\notag\\
\left(\gen{K}_1^\mu + \gen{K}_2^\mu\right) \frac{1}{x_{12}^2+z_1 \bar{\m}_2} &= \frac{2i \brk*{(1-\Delta_1) x_1^\mu + (1-\Delta_2)x_2^\mu)}}{x_{12}^2+\m_1 \bar{\m}_2},
\end{align}
which can clearly be made to vanish by a judicious choice of scaling dimensions $\Delta_i$. 
Since tree-level diagrams are simple products of single propagators, and since all the conformal generators are first order differential operators, the covariance factor of the whole amplitude is given by the sum of all single-propagator covariance factors. Again, for any diagram there is a choice of scaling dimensions for which the conformal generators annihilate the diagram. To lift this invariance to the level of amplitudes, note that due to the appearance of a three-point vertex in the Lagrangian \eqref{eq:LagrFishSpon},
\begin{equation}
\mathcal{L}_\text{SBF} = \mathcal{L}_\text{F} 
+ N_\text{c} \m_a \bar \m_b {\bar\phitwo^a}{}_b {\phitwo^b}_a 
+ N_\text{c} \xi \brk*{\bar \m_a
{X^a}{}_b {Z^{b}}{}_c{{\bar X}^c}{}_a
+ \m_a{\bar X^{a}}{}_b 
{\bar Z^{b}}{}_c {X^{c}}{}_a},
\end{equation}
the one-to-one correspondence between amplitudes and diagrams breaks down in the spontaneously broken phase. In particular, there can be different diagrams contributing to the same color-ordered tree-level amplitude, which arise from breaking up a four-point vertex according to 
\begin{align}
\includegraphicsbox{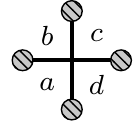}
+
\includegraphicsbox{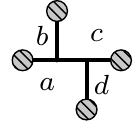}
+
\includegraphicsbox{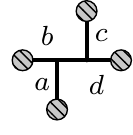}
\sim
 1 + \frac{\m_a \bar{\m}_c}{x_{ac}^2 + \m_a \bar{\m}_c} + \frac{\m_d \bar{\m}_b}{x_{db}^2 + \m_d \bar{\m}_b}.
\end{align}
All of these contributions have the same covariance behavior under the action of $\gen{D}$ and $\gen{K}^\mu$. Therefore all diagrams contributing to a given tree-level amplitude are rendered invariant by the same choice of $\Delta_i$, which implies that all tree-level amplitudes of the theory are dual conformal.

%%%%%%
Note that for the VEV scenario distinguished in \Secref{sec:PlanarAndVEV}, all integrals contributing to loop-level amplitudes in the SBF theory are identical to the ones appearing in the massless fishnet theory (up to massive tree-level pieces attached to massless loops). In fact, due to the product propagator and the necessity of massless internal regions to avoid pseudo-planar diagrams, all propagators that bound internal regions are massless. 
Hence, loop amplitudes in this theory are essentially the same as in the massless theory and thus feature the same symmetries, i.e.\ level-zero and level-one dual conformal alias Yangian symmetry (see \Appref{sec:DualConfSBFLoops} for an example on the level-zero symmetry). These symmetries, however,  do not provide us with any information about massive Feynman integrals, since the latter do simply not contribute to the planar limit described above.

%%%%%%%%%%%%%%%

\subsection{Massive Fishnet Theory}
In the massive fishnet theory, the situation differs significantly. Since the theory features difference-mass propagators, as does $\mathcal{N}=4$ SYM theory on the Coulomb branch, color-ordered amplitudes are invariant under the action of the same dual conformal generators \cite{Alday:2009zm}, i.e.\ 
\begin{align}
\gen{D}_i &= -i \left(x_i^\mu \frac{\partial}{\partial x_i^\mu} + m_i \frac{\partial}{\partial m_i} + \Delta_i\right),\notag\\
\gen{K}_{i}^\mu &= -2ix_i^\mu\left(x_i^\nu  \frac{\partial}{\partial x_i^\nu}  + m_i\frac{\partial}{\partial m_i} + \Delta_i\right) +i (x_i^2 +m_i^2)\frac{\partial}{\partial x_{i\mu}}.
\label{eq:DualConformalGenerators}
\end{align}
%

%%%%%%%

\paragraph{Level-Zero Dual Conformal Symmetry.} 

Again, level-zero dual conformal symmetry of any given tree-level diagram in the massless fishnet theory follows from covariance of the propagator:
\begin{align}
\left(\gen{D}_1 + \gen{D}_2\right) \frac{1}{x_{12}^2+ (m_1 -m_2)^2} &= \frac{i(2-\Delta_1 - \Delta_2)}{x_{12}^2 + (m_1 -m_2)^2},
\notag\\
\left(\gen{K}_1^\mu + \gen{K}_2^\mu\right) \frac{1}{x_{12}^2+(m_1 -m_2)^2} &= \frac{2i \brk*{(1-\Delta_1) x_1^\mu + (1-\Delta_2)x_2^\mu)}}{x_{12}^2+(m_1 -m_2)^2}.
\end{align}
Loop-level fishnet diagrams with massless internal regions are again invariant since the massless dual conformal generator is essentially a total derivative and can be inserted in loop integrals. Since there are no three-point vertices in the massive fishnet theory, there is a one-to-one mapping between color-ordered scattering amplitudes and Feynman diagrams, and this symmetry immediately carries over to the amplitude level.

Note that in contrast to the SBF theory discussed in \Secref{sec:SymsSBF}, due to the difference-mass propagators the internal massless regions do not render the entire loop integral massless. The diagrams surviving in the planar limit are fishnet diagrams with massless internal lines in the `bulk' of the diagram and a surrounding massive line. Some low-loop examples are given in \Figref{Fig:massiveFishnets}.
Therefore the symmetries of this theory are inherited by individual massive Feynman integrals. 
\begin{figure}
\begin{center}
a) \includegraphicsbox{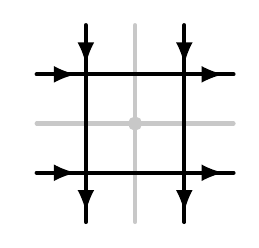}
b) \includegraphicsbox{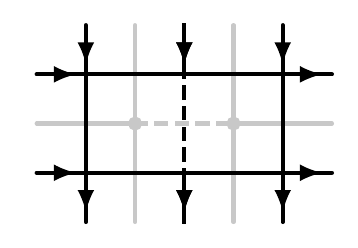}
c) \includegraphicsbox{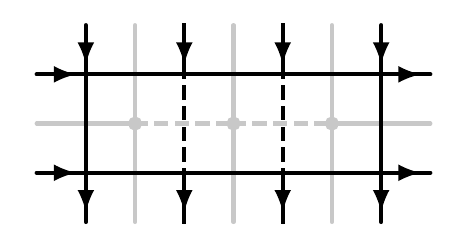}
\par
d) \includegraphicsbox{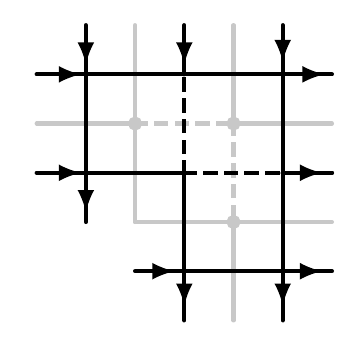}
\quad
e) \includegraphicsbox{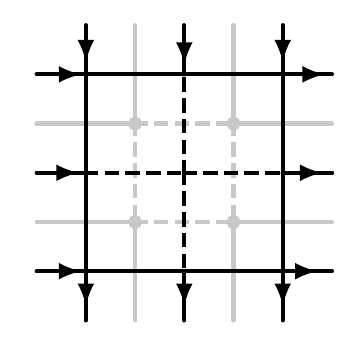}
\end{center}
\caption{Momentum-space Feynman diagrams (black) in one-to-one correspondence with one-particle-irreducible color-ordered amplitudes in the planar limit of the massive fishnet theory. Solid (dashed) lines correspond to massive (massless) propagators. Gray lines indicate the dual graph corresponding to the same Feynman integral expressed in terms of region momenta.}
\label{Fig:massiveFishnets}
\end{figure}

Note that the above generators can be interpreted as generating a subgroup of five-dimensional dual conformal symmetry where the mass variable $m_i$ plays the role of the fifth component of the vector $x_i^\mu$, i.e.\ $x_i^{5}=m_i$.  
Hence, the massive level-zero generators given above are obtained from the massless generators by adding this extra-dimensional component:
\begin{align}
\gen{D} &\to \gen{D}+\gen{D}_\fd,
&&\gen{D}_\fd=-i\sum_{i=1}^n x_i^5 \frac{\partial}{\partial x_i^5},
\\
\gen{K}^\mu &\to \gen{K}^\mu + \gen{K}^\mu_\fd,
&&
 \gen{K}^\mu_\fd=\sum_{i=1}^n \brk*{-2ix_i^\mu x_i^5\frac{\partial}{\partial x_i^5}+i(x_i^5)^2\frac{\partial}{\partial x_{i\mu}}}.
\end{align}
Interestingly, tree-level diagrams are invariant under an even larger symmetry algebra. To be explicit, the algebra can be extended to the full five-dimensional dual conformal algebra by the following six extra operators:
\begin{align}
\gen{P}_i^5 &= -i \frac{\partial}{\partial m_i},\quad\qquad\qquad \gen{L}_i^{5\nu} = i \left(m_i \frac{\partial}{\partial x_{i\nu}} - x_i^\nu \frac{\partial}{\partial m_i}\right),\notag\\
\gen{K}_i^5 &= 2i m_i \left(x_i^\nu \frac{\partial}{\partial x_i^\nu} + m_i \frac{\partial}{\partial m_i} + \Delta_i\right) - i (x_i^2 + m_i^2)\frac{\partial}{\partial m_i}.
\label{eq:ExtraGenerators}
\end{align}
Here again $m_i=x_i^5$.
It is easy to see that $\gen{P}^5$ is a symmetry of tree-level amplitudes, since it is the generator of translations in the mass parameters. Invariance under the other generators then follows from commuting $\gen{P}^5$ with generators of the four-dimensional dual conformal algebra. However, this extra tree-level symmetry is broken at loop level, since the planar limit described above sets all internal region masses to zero, breaking the invariance under~$\gen{P}^5$.

\paragraph{Level-One Dual Conformal Symmetry at Tree Level and One Loop.} 
As for the level-zero symmetry, we also consider an extra-dimensional contribution to the level-one generators:
\begin{equation}
\widehat{\gen{J}}^A_{(y)}= \widehat{\gen{J}}^A +y\, \widehat{\gen{J}}^A_\fd.
\label{eq:JhatExtra}
\end{equation}
Here the term proportional to the parameter $y$ arises from extending the sum over internal spacetime indices over $D=5$ instead of $D=4$ dimensions.
For the level-one momentum generator for instance, we have
\begin{equation}
\label{eq:PhatExtra}
\gen{\widehat P}^{ \mu}_\fd=\sfrac{i}{2}\sum_{j,k=1}^n\text{sign}(j-k)\gen{L}_j^{\mu5}\gen{P}_{k5} ,
\end{equation}
which is understood as a contribution originating from a sum over spacetime indices $\hat \nu$ ranging now over five dimensions:
\begin{equation}
\widehat{\gen{P}}^\mu_{(y=1)} = \frac{i}{2} \sum_{j,k=1}^n \text{sign}(j-k)\brk*{\gen{D}_j \gen{P}_k^\mu  + \gen{L}_j^{\mu\hat\nu} \gen{P}_{k\hat \nu}}  + \sum_{j=1}^n s_j \gen{P}_j^\mu.
\label{eq:ExPhat}
\end{equation}
It is useful to treat the extra generators \`a la \eqref{eq:PhatExtra} separately since for certain cases they annihilate Feynman integrals on their own and can thus be used to constrain these, cf.~\cite{Loebbert:2020hxk}.
The evaluation parameter $s_j$ entering the above level-one generator $\gen{\widehat J}_{(y)}^A$ can be constructed from the corresponding Feynman diagrams by the following rules adapted from \cite{Loebbert:2020hxk} (in the present paper we have $D=4$ and unit propagator weights):\footnote{Note that due to the dual-conformal symmetry of the amplitudes, one is actually free to shift all evaluation parameters by a common offset. For consistency, we still stick to the conventions of \cite{Loebbert:2020hxk}, where non-dual-conformal integrals were considered as well.} 
\begin{enumerate}
\item Start at point $x_1$ of the dual momentum graph and set $s_1 =(n-1)/2$.
\item To derive $s_j$, walk from point $1$ to point $j$ clockwise along the external boundary of the ($x$-space) graph and add the respective propagator weights, 
\item where an external propagator contributes $-1/2$ while an internal propagator contributes $(D-2)/2=1$.
\item All unconnected external points $x_j$ of the dual graph have $s_j=0$.
\end{enumerate}
To illustrate these rules, consider the ten-point amplitude
\begin{equation}
A(Z,\bar{X},\bar{X},\bar{Z},\bar{Z},\bar{Z},X,X,Z,Z)= \includegraphicsbox[scale=1]{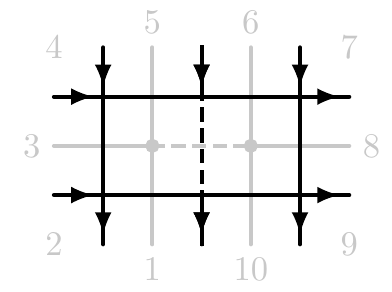}
.
\end{equation}
 The only contributing diagram is the two-loop double-box diagram in momentum space or the double-cross diagram in dual-momentum space, respectively. The evaluation parameters constructed from the above rules read
\begin{equation}
s_j=
\half(9,0,7,0,5,5,0,3,0,1)_j.
\end{equation}
Effectively, the 10-point level-one generator constructed in this way reduces to a 6-point operator that acts non-trivially only on the six external points of the dual (gray) Feynman graph, cf. \cite{Loebbert:2019vcj}.

\begin{figure}
\begin{center}
\includegraphicsbox{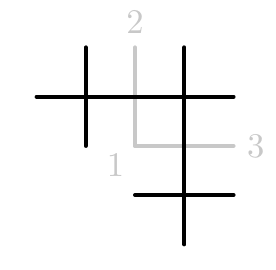}
$\quad \to \quad$
\includegraphicsbox{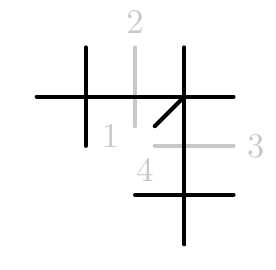}
\end{center}
\caption{On the left hand side we have an example of a massive tree-level fishnet diagram with multiple propagators adjacent to the same external region 1 (or the dual diagram with multiple propagators connected to the same external region momentum). At tree level and one loop such diagrams are annihilated by the level-one symmetry with extra-dimensional contribution $\widehat{\gen{J}}^A +\widehat{\gen{J}}^A_\fd$. At higher loops we introduce a point-splitting in region momentum space alias an additional leg in momentum space (right hand side), such that the diagram is annihilated by the generators $\widehat{\gen{J}}^A$ without extra term.}
\label{Fig:PointSplit}
\end{figure}

In general, the level-one generators $\gen{\widehat J}^A_{(y=1)}$, including the extra-dimensional contribution, annihilate color-ordered amplitudes $A_n$ in the massive fishnet theory at tree level and one loop:
\begin{align}
\gen{\widehat J}^A_{(y=1)} A_n = 0.
\end{align}
We find that in order
to have the full symmetry, some of the amplitudes%
\footnote{The corresponding Feynman diagrams were not considered in \cite{Loebbert:2020hxk}.} require the inclusion of the extra-dimensional contributions to the generators in \eqref{eq:JhatExtra}, while for others it is optional and represents a symmetry on its own, cf.~\cite{Loebbert:2020hxk}.  
The diagrams which require the 5D sum, i.e.\ $y=1$, are those that feature external regions with more than one adjacent propagator, see \Figref{Fig:PointSplit}. In terms of the dual graph, this special class of diagrams is classified by the fact that external region momenta are attached to multiple propagators. 
Moreover, while the level-zero extra symmetries \eqref{eq:ExtraGenerators} are broken by loop corrections, it was found in \cite{Loebbert:2020hxk} that the corresponding extensions (e.g.\ \eqref{eq:PhatExtra}) to the level-one generators  independently annihilate all one-particle-irreducible one-loop diagrams.
Going beyond the scope of \cite{Loebbert:2020hxk}, we explicitly checked that these also annihilate tree amplitudes as well as one-loop amplitudes with additional tree-level parts glued to them.

\paragraph{Level-One Dual Conformal Symmetry at Higher Loops.} 
Starting from two loops the extra generators $\gen{\widehat J}_\fd^A$ do not furnish symmetries.
For most loop amplitudes the statement 
\begin{equation}
\gen{\widehat J}^A A_n = 0
\end{equation} 
is implied (up to two loops) or conjectured (beyond two loops) by the findings of \cite{Loebbert:2020hxk}. This statement, however, does not hold for those graphs that feature external regions with more than one adjacent propagator, see \Figref{Fig:PointSplit}.
 To turn the massive fishnet amplitudes into invariants under the generators with $y=0$, we introduce a point-splitting procedure: If coincident external points are treated as separate, the generator $\gen{\widehat J}^A$ annihilates the amplitude. In momentum space this corresponds to introducing extra external legs between adjacent propagators. This modified amplitude then features the level-one symmetry with 4D sums over spacetime indices, and the actual massive fishnet amplitude can be recovered by taking soft limits in the extra legs. An example of this procedure is shown in \Figref{Fig:PointSplit}. 
There is also an interest in not including extra-dimensional contributions to the level-one generators, i.e.\ in setting $y=0$, since this simplifies the resulting operator when translated to momentum space (cf.\  \eqref{eq:Lev1InMom} below).

%%%

\paragraph{Momentum Space Interpretation of Level-One Symmetry.}
In \cite{Loebbert:2020hxk} the level-one symmetry $\gen{\widehat P}^\mu$ was also translated back to momentum space, where it leads to invariance under the action of a massive conformal algebra generated by the ordinary momentum and Lorentz generators 
\begin{equation}
\gen{\bar{P}}_i^\mu = p_i^\mu, 
\qquad\qquad
 \gen{\bar{L}}_i^{\mu\nu} =  p_i^\mu \frac{\partial}{\partial p_{i\nu}} - p_i^\nu \frac{\partial}{\partial p_{i\mu}},
\end{equation}
as well as the extended conformal generators
\begin{align}
\bar{\gen{D}}_j &= p_{j\nu} \frac{\partial}{\partial p_{j\nu}} + \frac{1}{2}\brk*{m_j \frac{\partial}{\partial m_j} + m_{j+1} \frac{\partial}{\partial m_{j+1}}} + \bar{\Delta}_j,
\notag \\
  \bar{\gen{K}}^\mu_j &= p_j^\mu \frac{\partial^2}{\partial p_j^\nu \partial p_{j\nu}} - 2 \bar{\gen{D}}_j \frac{\partial}{\partial p_{j\mu}}.
  \label{eq:Lev1InMom}
  \end{align}
In the cases that require the point-splitting procedure described above, physical legs have $\bar{\Delta_j}=1$, whereas the non-physical extra legs have $\bar{\Delta_j}=0$.
%%%%%%%%%%%%%%%%%%%%%%%%%%%%%%%%%%%%%%%%%%%%%%%%%%%%%%%%%%%%%%%%%%%%%%%%%%%

\section{Conformal Soft Theorems in the SBF Theory}
\label{sec:Soft}
%%%%%%%%%%%%%%%%%%%%%%%%%%%%%%%%%%%%%%%%%%%%%%%%%%%%%%%%%%%%%%%%%%%%%%%%%%%
In the previous section we have discussed the status of dual conformal symmetry in both the SBF theory as well as the massive fishnet theory, and of ordinary conformal symmetry in the latter theory. To complete the discussion, let us demonstrate the imprints of ordinary conformal symmetry on amplitudes in the SBF theory. 
It is well known that in theories with spontaneously broken conformal symmetry the conformal generators control the behavior of scattering amplitudes with soft Goldstone bosons
\cite{Boels:2015pta,DiVecchia:2015jaq} (see also \cite{Low:2001bw,Huang:2015sla,Luo:2015tat,Bianchi:2016viy,Rodina:2018pcb}). Due to the limited amount of diagrams, the SBF theory furnishes a pedagogical example of this mechanism. For the moment we restrict ourselves to tree-level amplitudes.

%%%%%%
\paragraph{Reparametrization of SBF Theory.}

To properly exhibit the soft behavior, the action of the SBF theory must be parametrized in terms of the Goldstone boson mode. 
The appropriate parametrization of the field $\phione$ is given by
\begin{align}
{\phione^a}_b = \left[\left(\frac{\m_a}{\xi} + \frac{1}{\sqrt{2}}\sigma(x) \frac{\m_a}{\m}\right)\delta^a_b + {\varphi^a}_b\right]e^{ i\sfrac{\xi}{\sqrt{2}\m_a} \rho(x)},
\end{align}
where $\sigma$ is the Goldstone boson related to spontaneously broken scale invariance (dilaton), whereas $\rho$ is the Goldstone boson related to the broken $\grp{U}(1)$ symmetry. We furthermore define
\begin{align}
\m^2 = \sum_{a=1}^{N_\text{c}} \m_a \bar \m_a.
\end{align}
Note that $\varphi$ takes values only in the subset of $\alg{su}(N)$ which is orthogonal to $\m_a\delta^a_b/\m $ with respect to the scalar product $\langle \,\cdot\, ,\,\cdot\,\rangle = \text{Tr}(\,\cdot\, \,\cdot\,)$.
We then have
\begin{align}
-N_c\text{Tr}(\partial_\mu \bar\phione \partial^\mu \phione) = \mathcal{L}_\text{kin} + \mathcal{L}_{\text{int}, \rho},
\end{align}
where we have defined the kinetic part of the Lagrangian as
\begin{align}
\mathcal{L}_\text{kin} =-N_c\brk*{\sfrac{1}{2} \partial_\mu \sigma \partial^\mu \sigma + \sfrac{1}{2} \partial_\mu \rho \partial^\mu \rho +\text{Tr}(\partial_\mu \bar\varphi \partial^\mu \varphi)},
\end{align}
as well as interaction part
\begin{align}
\mathcal{L}_{\text{int},\rho} =&\ -i \frac{\xi}{\sqrt{2}\m}\partial_\mu \rho \text{Tr}(\varphi \partial^\mu \bar\varphi - \partial^\mu \varphi \bar\varphi) 
- \frac{\xi}{\sqrt{2}\m} \partial_\mu \rho \partial^\mu \rho \sigma\\
&\  -  \frac{\xi^2}{4\m^2} \partial_\mu \rho \partial^\mu \rho \sigma^2 
-  \frac{\xi^2}{2\m^2}\partial_\mu \rho \partial^\mu \rho \text{Tr}(\varphi \bar\varphi).
\end{align}
Upon expanding $\phione$ in the above way, the original interaction term becomes
\begin{align}
 \xi^2 \text{Tr}(\bar\phitwo \bar\phione \phitwo \phione) 
=
&
  \m_a \bar\m_b {\bar\phitwo^a}_b {\phitwo^b}_a+
\xi \left(\bar \m_b{\bar\phitwo^a}{}_b {\phitwo^b}_c{\varphi^c}_a  +
\m_a {\bar\phitwo^a}_b {\bar\varphi^b}{}_c {\phitwo^c}_a \right)\notag\\
&\ +  \frac{\xi^2}{\sqrt{2}\m} \sigma \left(\bar\m_b{\bar\phitwo^a}{}_b {\phitwo^b}_c {\varphi^c}_a + 
 \m_b {\phitwo^a}_b {\bar\phitwo^b}{}_c {\bar \varphi^c}{}_a\right) +
 \sqrt{2} \frac{\xi}{\m} \sigma  \m_a \bar \m_b {\bar\phitwo^a}{}_b{\phitwo^b}_a\notag\\
&\ + \frac{\xi^2}{2\m^2} \sigma^2  \m_a \bar \m_b {\bar\phitwo^a}{}_b{\phitwo^b}_a + 
 \xi^2 {\bar\phitwo^a}_b {\bar\varphi^b}{}_c {\phitwo^c}_d {\varphi^d}_a.
\end{align}
%%%%%%

\paragraph{Color Ordering.}

Amplitudes in this theory can be color-ordered only with respect to their external $\phitwo$ and $\varphi$ legs. As an example, consider the six-point amplitude 
\begin{align}
\mathcal{A}_6 = \langle {\bar\phitwo(p_1)^{a_1}}_{b_1} {\bar\phitwo(p_2)^{a_2}}_{b_2}  {\varphi(p_3)^{a_3}}_{b_3}  {\phitwo(p_4)^{a_4}}_{b_4} {\phitwo(p_5)^{a_5}}_{b_5} \sigma(p_6)\rangle.
\end{align} 
It decomposes into different color structures according to
\begin{align}
\mathcal{A}_6 = \sum_{s \in S_5/Z_5} \delta_{a_{s(2)}}^{b_{s(1)}}\delta_{a_{s(3)}}^{b_{s(2)}}\delta_{a_{s(4)}}^{b_{s(3)}}\delta_{a_{s(5)}}^{b_{s(4)}} \delta_{a_{s(1)}}^{b_{s(4)}} A_6(\Phi_{s(1)},\Phi_{s(2)},\Phi_{s(3)},\Phi_{s(4)}, \Phi_{s(5)}|\sigma(p_6)),
\end{align}
where $\vec \Phi$ is a vector containing all colored external fields in the amplitude, including their full kinematic, flavor and color information, i.e.\
\begin{align}
\vec\Phi = \left({\bar\phitwo(p_1)^{a_1}}_{b_1}, {\bar\phitwo(p_2)^{a_2}}_{b_2},  {\varphi(p_3)^{a_3}}_{b_3},  {\phitwo(p_4)^{a_4}}_{b_4}, {\phitwo(p_5)^{a_5}}_{b_5}\right).
\end{align}
In this example, only a subset of the color structures is non-vanishing. Explicitly, the non-zero color-ordered tree-level amplitudes read
\begin{align}
&A_6\brk!{{\bar\phitwo(p_1)^{a_1}}_{b_1}, {\bar\phitwo(p_2)^{a_2}}_{b_2},  {\varphi(p_3)^{a_3}}_{b_3},  {\phitwo(p_4)^{a_4}}_{b_4}, {\phitwo(p_5)^{a_5}}_{b_5}| \sigma(p_6)} \notag\\&=  \includegraphicsbox[scale=0.8]{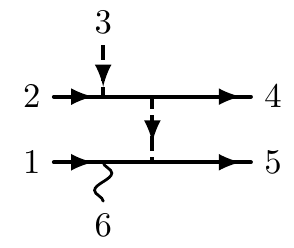}+\includegraphicsbox[scale=0.8]{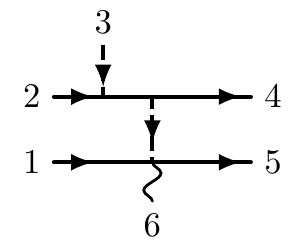}+\includegraphicsbox[scale=0.8]{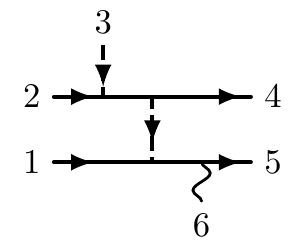}+ (5\text{ others})\notag\\
&\ + \includegraphicsbox[scale=0.8]{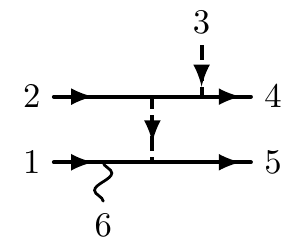}+\includegraphicsbox[scale=0.8]{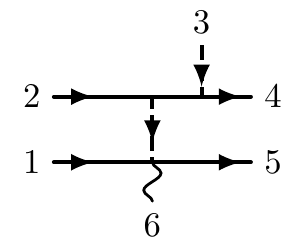}+\includegraphicsbox[scale=0.8]{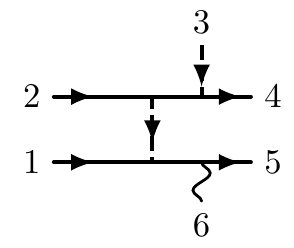}+(5\text{ others})\notag\\
&\ + \includegraphicsbox[scale=0.8]{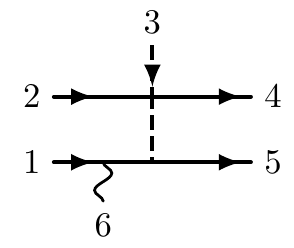}+\includegraphicsbox[scale=0.8]{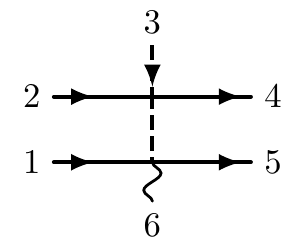}+\includegraphicsbox[scale=0.8]{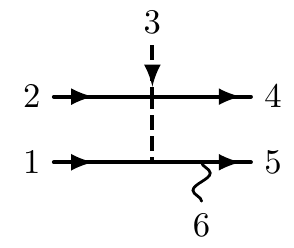}+(2\text{ others})\notag,
\end{align}
as well as
\begin{align}
&A_6\brk!{{\bar\phitwo(p_1)^{a_1}}_{b_1},{\varphi(p_3)^{a_3}}_{b_3}, {\bar\phitwo(p_2)^{a_2}}_{b_2},  {\phitwo(p_4)^{a_4}}_{b_4},   {\phitwo(p_5)^{a_5}}_{b_5}| \sigma(p_6)}\\&= \includegraphicsbox[scale=0.8]{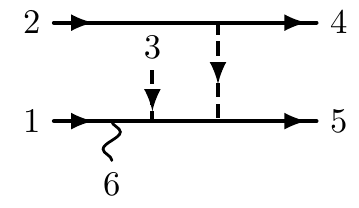}+\includegraphicsbox[scale=0.8]{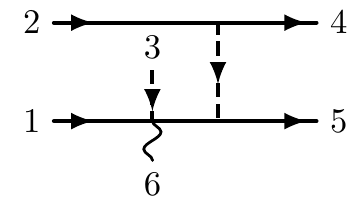}+\includegraphicsbox[scale=0.8]{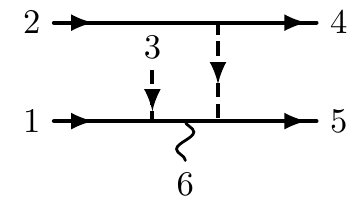} + (5\text{ others}),
\end{align}
and the amplitudes that are related to the above by the permutations of the external $\phitwo$ and $\bar\phitwo$ fields, respectively. Obviously, partial amplitudes with different color-orderings may have different contributing graph topologies. Moreover the  Goldstone boson carries no color and does therefore not respect the color-ordering but can attach to any massive line of the diagrams.
%%%%%%%%%%%%%%%%%%%%%%%%%%%%%%%%%%%%%%%%%%%%%%%%%%%%%%%%%%%%%%%%%%%%%%%%%%%
\paragraph{Soft Theorems.}

Before the spontaneous symmetry breaking, the fishnet amplitudes featured a conformal symmetry,%
\footnote{We ignore potential conformal anomalies, cf.\ \cite{Bargheer:2009qu,Sever:2009aa,Bargheer:2011mm,Bargheer:2012cp,Bianchi:2012cq,Chicherin:2017bxc}.}
that is, the tree amplitudes were invariant under the tensor product representation
\begin{equation}
\gen{\bar{J}}^A =\sum_{k=1}^n \gen{\bar{J}}_k^A,
\end{equation}
where the index $A$ runs over the set of conformal generators:
\begin{align}
\gen{\bar{P}}_i^\mu &= p_i^\mu, 
&\gen{\bar{K}}_i^\mu &= p_i^\mu \frac{\partial^2}{\partial p_i^\nu \partial p_{i\nu}}-2 \left(p_i^\nu \frac{\partial}{\partial p_i^\nu} +  \Delta_i \right)\frac{\partial}{\partial p_{i\mu}},
\notag\\
\gen{\bar{D}}_i &=p_i^\mu \frac{\partial}{\partial p_i^\mu} + \Delta_i, 
& \gen{\bar{L}}_i^{\mu\nu} &=  p_i^\mu \frac{\partial}{\partial p_{i\nu}} - p_i^\nu \frac{\partial}{\partial p_{i\mu}}.
\label{eq:pspacegens}
\end{align}
These generators satisfy the commutation relations given in \Appref{app:ConformalCommutators}.
Note that on scalar amplitudes where $\Delta_i=1$, $\gen{\bar{K}}$ commutes with the on-shell constraint $p^2=0$, which is not true in general, cf.\ \cite{Loebbert:2018xce}.
In the broken phase, this leads to 
the soft theorem for the color-ordered amplitudes 
\cite{Boels:2015pta,DiVecchia:2015jaq,Luo:2015tat}
\begin{align}
\lim_{\tau\rightarrow 0} A_n(...|\sigma(\tau p_n)) = \frac{\xi}{\sqrt{2}\m}\brk[s]*{\frac{1}{\tau} \gen{S}_m^{(-1)} + \gen{S}_m^{(0)} + \gen{S}^{(0)} + \tau (\gen{S}_m^{(1)}+\gen{S}^{(1)})}A_{n-1}(...) + \mathcal{O}(\tau^2),\label{eq:softtheorem}
\end{align}
where we have rescaled $p_n\to \tau p_n$ such that $\tau\to 0$ corresponds to taking the momentum $p_n$ soft. Moreover, we have defined
\begin{equation}
\gen{S}_m^{(-1)} = \sum_{k=1}^{n-1} \frac{m_k^2}{p_k\cdot p_n} ,
\qquad
\gen{S}_m^{(0)} = \sum_{k=1}^{n-1} \frac{m_k^2}{p_n\cdot p_k} p_n\cdot \frac{\partial}{\partial p_k},
\qquad
\gen{S}_m^{(1)} = \sum_{k=1}^{n-1} \frac{1}{2} \frac{m_k^2}{p_n\cdot p_k} p_n^\mu p_n^\nu \frac{\partial^2}{\partial p_k^\mu \partial p_k^\nu} ,
\end{equation}
as well as using the generators $\gen{\bar{D}}$ and $\gen{\bar{K}}$ of \eqref{eq:pspacegens}:
\begin{equation}
\gen{S}^{(0)} = -\,\gen{\bar{D}}+4,
\qquad
\gen{S}^{(1)} = \half\,p_n\cdot \gen{\bar{K}}.
\end{equation}
Here the $m_k$ refer to the physical masses of the external legs $k$, i.e.\ we have
\begin{align}
 m({{\varphi}^a}_b)&=0 ,&   m({{\phitwo}^a}_b) &= \m_a \bar \m_b.
\end{align}
Since the soft operators only depend on the external kinematic data and not on the ordering of the legs, they are the same for differently color-ordered partial amplitudes contributing to the same full amplitude. Therefore, the soft theorem also holds at the level of the full amplitudes.
%%%%%%%
\paragraph{Examples.}

The soft theorem applies to amplitudes in the SBF theory. We checked it explicitly for the following examples:%
\footnote{Listed here are the $n-1$ point amplitudes on the right-hand side of the soft-theorem.}
\begin{itemize}
\item[a/b)] the three-point amplitude with one massless and two massive legs (the T-amplitude),
\item[c/d)] the four-point amplitudes with two massless and two massive legs (the cross and the table amplitude),
\item[e)] the four-point amplitude with four massive legs (the H-amplitude),
\item[f)] the four-point amplitude with one massless and two massive legs as well as one Goldstone boson (thus, in this case the soft limit of a two-Goldstone-boson amplitude is taken),
\item[g)] the six-point amplitude with six massive legs,
\item[h)] the six-point amplitude with two massive and four massless legs (the double-cross amplitude).
\end{itemize}
\begin{table}
\begin{center}
\begin{tabular}{c l c l c l c l }
a) & \includegraphicsbox[scale=0.8]{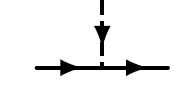}
&
b) & \includegraphicsbox[scale=0.8]{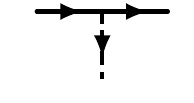}
&
c) & \includegraphicsbox[scale=0.8]{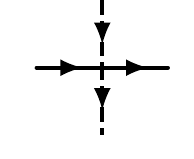} +\dots
&
d) & \includegraphicsbox[scale=0.8]{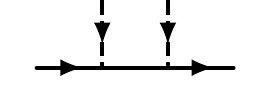}+\dots
\\
e) & \includegraphicsbox[scale=0.8]{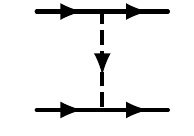}
&
f) &\includegraphicsbox[scale=0.8]{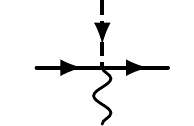}+\dots
&
g) &\includegraphicsbox[scale=0.8]{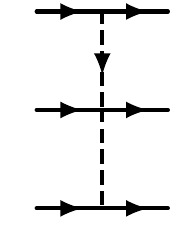}+\dots
&
h) &\includegraphicsbox[scale=0.8]{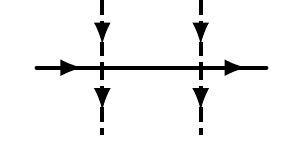}
\end{tabular}
\end{center}
\caption{List of amplitudes for which the soft theorem has been checked. In particular, the list contains all tree-level amplitudes with four external legs or less. In the cases where multiple diagrams contribute, we have picked a single representative diagram. All other diagrams can be constructed by permuting external legs and splitting up internal four-point vertices.}
\label{Fig:massiveFishnets}
\end{table}
As an explicit example, consider the second to last case. Up to cyclic permutations, there are three non-vanishing topologies contributing to color-ordered amplitudes, i.e.
\begin{align}
A_6(\phitwo_1, \phitwo_2,\phitwo_3,\bar{\phitwo}_4,\bar{\phitwo}_5,\bar{\phitwo}_6) &=\includegraphicsbox[scale=0.8]{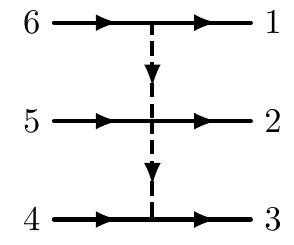}+\includegraphicsbox[scale=0.8]{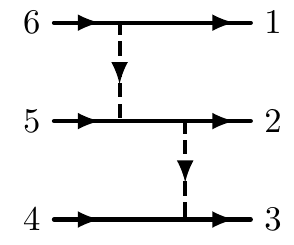}+\includegraphicsbox[scale=0.8]{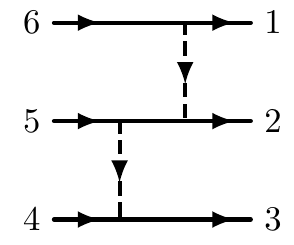},\notag\\
A_6(\phitwo_1, \phitwo_2,\bar{\phitwo}_4,\phitwo_3,\bar{\phitwo}_5,\bar{\phitwo}_6) &= \includegraphicsbox[scale=0.8]{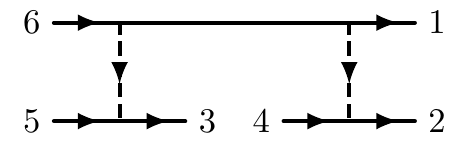},\notag\\
A_6(\phitwo_1, \phitwo_2,\bar{\phitwo}_4,\bar{\phitwo}_5,\phitwo_3,\bar{\phitwo}_6) &= \includegraphicsbox[scale=0.8]{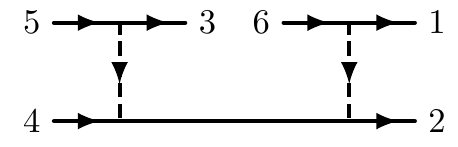}.\notag\\
\end{align}
Acting on these amplitudes with the soft operators in \eqref{eq:softtheorem}, one recovers the soft limit expansion of the amplitude with an additional Goldstone boson attached to the above diagrams at all possible locations.

%%%%%%%%%%%%%%%%%%%%%%%%%%%%%%%%%%%%%%%%%%%%%%%%%%%%%%%%%%%%%%%%%%%%%%%%%%%
\section{Conclusions and Outlook}
\label{sec:Conclusions}

The most important results of this paper are summarized as follows.
 Generalizing the derivation of the massless fishnet theory of \cite{Gurdogan:2015csr}, in \Secref{sec:MFLimit} we defined a massive fishnet theory as a particular double-scaling limit of $\superN=4$ SYM theory on the Coulomb branch. In the resulting theory, planar scattering amplitudes are conjecturally  in one-to-one correspondence%
\footnote{As in the massless case, a general proof of this one-to-one correspondence is still lacking.} with individual massive Feynman integrals of fishnet structure. The theory is  integrable in the sense that these off-shell amplitudes are annihilated by a massive extension of Yangian symmetry as recently shown in \cite{Loebbert:2020hxk}.%
\footnote{Here we emphasize the word `off-shell' since potential subtleties in the on-shell case have not been investigated yet, cf.\ \cite{Bargheer:2009qu,Sever:2009aa,Bargheer:2011mm,Bargheer:2012cp,Bianchi:2012cq,Chicherin:2017bxc}.} Hence, the situation is very similar to the massless bi-scalar fishnet theory which justifies the name \emph{massive fishnet theory} for the theory defined by the Lagrangian
\begin{align}
\Lagr_\text{MF}=
&N_\text{c} \tr\brk*{
-\partial_\mu \bar X \partial^\mu X 
-
\partial_\mu \bar Z \partial^\mu Z
+\xi^2 \bar X\bar Z X Z
}
\nonumber\\
&-  N_\text{c}(m_a - m_b)^2 {X^a}_b{\bar X^b}_a
- N_\text{c}(m_a - m_b)^2 {Z^a}_b{\bar Z^b}_a.
\end{align} 
We have contrasted this massive fishnet theory with the situation of spontaneous symmetry breaking (SBF theory) in the massless fishnet theory. A crucial difference lies in how the masses enter into  propagators. While the above massive fishnet theory has massive propagators of difference form, spontaneous symmetry breaking leads to product-mass propagators. As a consequence, for the given VEV configuration, Feynman integrals with massive propagators drop out of scattering amplitudes in the planar limit of the latter theory. Moreover, it is not clear if the notion of Yangian symmetry can be extended to Feynman integrals with product-mass propagators. 
In \Secref{sec:Soft} we have explicitly demonstrated that the SBF theory furnishes a pedagogical example where scattering amplitudes obey conformal soft theorems which are controlled by the generators of the (ordinary) conformal algebra. It would certainly be interesting to extend these tests to higher loop orders  and to see how the loop-conditions discussed in \cite{Karananas:2019fox} enter.
\bigskip

In the massless case one can argue that Feynman integrals inherit their integrability from $\superN=4$ SYM theory via the fishnet theory. Here we followed the reverse logic and based our motivation on the Yangian symmetry of massive Feynman integrals, which we now understand as planar off-shell amplitudes of the massive fishnet theory defined through the Lagrangian \eqref{eq:FullLagrBiI}.
The derivation of this massive fishnet theory from Higgsed $\superN=4$ SYM theory raises the question whether integrability can also be formulated in the latter. 
In particular, it might be interesting to take a closer look at Coulomb-branch scattering amplitudes with regard to the massive Yangian symmetry discussed here. As a complementary approach one may wonder whether the methods of \cite{Beisert:2017pnr,Beisert:2018zxs,Beisert:2018ijg} can be extended to show a Yangian invariance of the action of Coulomb-branch $\superN=4$ SYM theory. That would close the remaining gap in the argument to interpret integrability of the massive fishnet theory as being inherited from a massive integrable mother theory. As a third approach into this direction  it should be very enlightening to see if the holographic dual of the massless fishnet theory found in \cite{Gromov:2019aku,Gromov:2019bsj,Gromov:2019jfh} can be extended to the massive case. This would shed light on the interpretation of our findings in the context of the AdS/CFT duality.
\bigskip

It is natural to ask for the properties and potential integrability of the zoo of theories obtained from other double-scaling limits, e.g.\ the massive tri-scalar theory discussed at the end of \Secref{sec:MFLimit}. This tri-scalar theory shows that different field contents than in the massless case are reachable in the massive situation (cf.\ \cite{Caetano:2016ydc} for the massless case).
A next step would be a systematic investigation of the symmetry properties of the tri-scalar theory and to see whether it shows similar integrability properties as the massive fishnet theory. Moreover, one might expect that the tri-scalar theory features a soft theorem similar to the one discussed in \Secref{sec:Soft} for the SBF theory.
\bigskip

At the more technical level we note that in the derivation of the massive fishnet theory in \Secref{sec:MFLimit} we have ignored double-trace counter-terms at the intermediate step of the massless gamma-deformed theory \cite{Dymarsky:2005uh,Pomoni:2008de,Fokken:2013aea,Fokken:2014soa}. Similar terms have later been added to the original definition of the massless fishnet theory \cite{Sieg:2016vap,Grabner:2017pgm}. It would be important to investigate whether some observables in the massive fishnet theory require such extra terms for a consistent treatment.
Furthermore, we note that the observation of a massive Yangian symmetry in \cite{Loebbert:2020hxk} actually applies to $D$ spacetime dimensions. It is thus natural to investigate massive fishnet theories in dimensions different from $4$.
Moreover, here we have focussed on symmetry properties of scattering amplitudes and Feynman integrals in order to single out a massive fishnet theory. It should be interesting to also look at other observables and to see if integrability facilitates their computation. 
\bigskip

Finally, we note that the difference-mass propagators discussed in this paper can be understood as an integrability-preserving regulator for on-shell/collinear divergences of the respective Feynman integrals. It has been known for more than ten years that this mass-regulator preserves the dual conformal part of the Yangian algebra \cite{Alday:2009zm}. With the observation of \cite{Loebbert:2020hxk} we now know that 
in fact this regulator preserves a massive extension of the whole Yangian. It is thus a natural next step to exploit this powerful symmetry for the computation of Feynman integrals with on-shell massless external particles as they arise from the massless limit of the diagrams discussed here.

%%%%%%%%%%%%%%%%%%%%%%%%%%%%%%%%%%%%%%%%%%%%%%%%%%%%%%%%%%%%%%%%%%%%%%%%%%
\paragraph{Acknowledgements.}
We are grateful to Dennis M\"uller and Hagen M\"unkler for collaborations on closely related projects and to Lance Dixon, Jan Plefka, Amit Sever and Congkao Wen for helpful discussions. We thank Volodya Kazakov, Dennis M\"uller and Hagen M\"unkler for useful remarks on the manuscript. Moreover, we thank 
Georgios Karananas,  Volodya Kazakov and Mikhail Shaposhnikov for sharing an update of their letter \cite{Karananas:2019fox} prior to publication.
The work of FL is funded by the Deutsche Forschungsgemeinschaft (DFG, German Research Foundation)–Projektnummer 363895012. JM is supported  by the International Max Planck Research School for Mathematical and Physical Aspects of Gravitation, Cosmology and Quantum Field Theory.

%%%%%%%%%%%%%%%%%%%%%%%%%%%%%%%%%%%%%%%%%%%%%%%%%%%%%%%%%%%%%%%%%%%%%%%%%%

\appendix
%%%%%%%%
\section{Details on Gamma-Deformation}
\label{sec:DetailsGammaDef}

The $\alg{u}(1)^3$ charges $q_j$ entering \eqref{eq:biPhase} read 
\begin{center}
\renewcommand{\arraystretch}{1.3}
\begin{tabular}{|c|c|c|c|c|c|c|c|c|}\hline
&$\psi_\alpha^1$&$\psi_\alpha^2$&$\psi_\alpha^3$&$\psi_\alpha^4$&$A_\mu$&$\phi_1$&$\phi_2$&$\phi_3$
\\\hline
$q_1$&$+\half$&$-\half$&$-\half$&$+\half$&$0$&$1$&$0$&$0$
\\
$q_2$&$-\half$&$+\half$&$-\half$&$+\half$&$0$&$0$&$1$&$0$
\\
$q_3$&$-\half$&$-\half$&$+\half$&$+\half$&$0$&$0$&$0$&$1$
\\\hline
\end{tabular}
\end{center}
\medskip
To facilitate the comparison of different versions of the Lagrangian of $\superN=4$ SYM theory and its gamma-deformation given in the literature, we note the identity
\begin{equation}
\quarter \tr \acomm{\phi_i^\dagger}{\phi^i}\acomm{\phi_j^\dagger}{\phi^j}
=\quarter\tr  \comm{\phi_i^\dagger}{\phi^i}\comm{\phi_j^\dagger}{\phi^j}
+\tr \phi_i^\dagger\phi_j^\dagger \phi^j\phi^i.
\end{equation}
%%%%%%%%
\section{Commutators of Conformal Generators}
\label{app:ConformalCommutators}

The conformal generators \eqref{eq:pspacegens} in momentum space satisfy the following commutation relations:
\begin{align}
[\gen{\bar{D}}, \gen{\bar{P}}^\mu] &=  \gen{\bar{P}}^\mu, 
& [\gen{\bar{D}}, \gen{\bar{K}}^\mu] &= - \gen{\bar{K}}^\mu,
\notag\\
[\gen{\bar{P}}^\mu, \gen{\bar{L}}^{\nu\rho}] &= \left(\eta^{\mu\nu} \gen{\bar{P}}^\rho - \eta^{\mu\rho} \gen{\bar{P}}^\nu\right),
& [\gen{\bar{K}}^\mu, \gen{\bar{L}}^{\nu\rho}] &= \left(\eta^{\mu\nu} \gen{\bar{K}}^\rho - \eta^{\mu\rho} \gen{\bar{K}}^\nu\right),
\notag\\
[\gen{\bar{K}}^\mu, \gen{\bar{P}}^\nu] &= -2\left(\eta^{\mu\nu} \gen{\bar{D}} - \gen{\bar{L}}^{\mu\nu}\right),
& [\gen{\bar{L}}^{\mu\nu}, \gen{\bar{L}}^{\rho\sigma}] &= \left(\eta^{\mu\sigma}\gen{\bar{L}}^{\rho\nu} + (\text{3 more})\right).&
\end{align}
All sets of dual conformal generators in \Secref{sec:Yangian} satisfy the commutation relations 
\begin{align}
[\gen{D}, \gen{{P}}^\mu] &=  i\gen{{P}}^\mu, 
& [\gen{{D}}, \gen{{K}}^\mu] &= - i\gen{{K}}^\mu,
\notag\\
[\gen{{P}}^\mu, \gen{{L}}^{\nu\rho}] &= i\left(\eta^{\mu\nu} \gen{{P}}^\rho - \eta^{\mu\rho} \gen{{P}}^\nu\right),
& [\gen{{K}}^\mu, \gen{{L}}^{\nu\rho}] &= i\left(\eta^{\mu\nu} \gen{{K}}^\rho - \eta^{\mu\rho} \gen{{K}}^\nu\right),
\notag\\
[\gen{{K}}^\mu, \gen{{P}}^\nu] &= 2i\left(\eta^{\mu\nu} \gen{{D}} - \gen{{L}}^{\mu\nu}\right),
& [\gen{{L}}^{\mu\nu}, \gen{{L}}^{\rho\sigma}] &= i\left(\eta^{\mu\sigma}\gen{{L}}^{\rho\nu} + (\text{3 more})\right).&
\end{align}
Note that one could adopt a normalization in which both algebras are identical. We refrain from doing so to be in line with the existing literature.
%%%%%%%%
\section{Dual Conformal Level-One Generators}
\label{app:Level1Generators}
Explicitly, the conformal level-one generators are given by
\begin{align}
\widehat{\gen{P}}^\mu &= \frac{i}{2} \sum_{j,k=1}^n \text{sign}(j-k)\brk*{\gen{D}_j \gen{P}_k^\mu  + \gen{L}_j^{\mu\nu} \gen{P}_{k\nu}}  + \sum_{j=1}^n s_j \gen{P}_j^\mu,
\notag\\
\widehat{\gen{L}}^{\mu\nu} &= \frac{i}{2} \sum_{j,k=1}^n \text{sign}(j-k)\brk*{\sfrac{1}{2}\brk{\gen{K}_j^\nu \gen{P}_k^\mu-\gen{K}_j^\mu \gen{P}_k^\nu} + \gen{L}_j^{\mu \rho} \gen{L}_{k\rho}{}^\nu} + \sum_{j=1}^n s_j \gen{L}_j^{\mu\nu},
\notag\\
\widehat{\gen{D}} &= \frac{i}{2} \sum_{j,k=1}^n \text{sign}(j-k)\frac{1}{2} \gen{K}_j^\mu \gen{P}_{k\mu} + \sum_{j=1}^n s_j \gen{D}_j,
\notag\\
\widehat{\gen{K}}^{\mu} &= \frac{i}{2} \sum_{j,k=1}^n \text{sign}(j-k)\brk*{\gen{K}^\mu_j \gen{D}_k + \gen{L}_j^{\mu\nu} \gen{K}_{k\nu}} + \sum_{j=1}^n s_j \gen{K}_j^\mu,
\label{eq:Level1Generators}
\end{align}
where sums of double indices run over $\mu,\nu,\rho \in \{1,2,3,4\}$.

%%%%%%%%%%

\section{Dual Conformal Symmetry of Loop Amplitudes in the SBF Theory}
\label{sec:DualConfSBFLoops}

Let us discuss the  particular example of the box diagram, i.e.\ the amplitude 
\begin{align}
A_8(\bar{X}\bar{X}\bar{Z}\bar{Z}XXZZ) =\includegraphicsbox[scale=1]{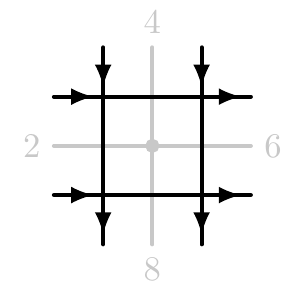}= \xi^8 \int \frac{\text{d}x_A^D}{x_{2A}^2 x_{4A}^2 x_{6A}^2 x_{8A}^2}.
\end{align}
Within the refined planar limit described in \secref{sec:PlanarAndVEV} the internal region is massless, leaving only a single diagram which is in fact identical to the corresponding massless amplitude. To prove that this amplitude (and all other loop amplitudes, even the ones that did not appear in the massless theory and have more than one contributing diagram) is covariant, note that the massless special conformal current $\gen{K}_A^\mu$ acting on the internal region is essentially a total derivative and can be inserted under the integral sign,
\begin{align}
(\gen{K}_A^\mu -2 i D x_A^\mu) f(x_A)= -i \frac{\partial}{\partial x_A^\nu}\brk*{ \brk*{2x_A^\mu x_A^\nu - x_A^2 \eta^{\mu\nu}} f(x_A)}.
\end{align}
Therefore, acting with the special conformal generator with  $\Delta_{2i}=1, \Delta_{2i+1}=0$ on the above amplitude, one finds
\begin{align}
\gen{K}^\mu A_8(\bar{X}\bar{X}\bar{Z}\bar{Z}XXZZ) &= \xi^8 \int \text{d}x_A^D \brk*{\gen{K}_1^\mu + \dots + \gen{K}_8^\mu + \gen{K}_A^\mu - 2i Dx_A^\mu}\frac{1}{x_{2A}^2x_{4A}^2 x_{6A}^2 x_{8A}^2}\notag\\
&= 2i \xi^8 \int \text{d}x_A^D \frac{(4-D) x_A^\mu }{x_{2A}^2x_{4A}^2 x_{6A}^2 x_{8A}^2},
\end{align}
which vanishes identically in $D=4$ dimensions. In the same way the symmetry can be proved to hold for any loop diagram with an arbitrary number of loops: by inserting a massless dual conformal generator for every internal region and using the covariant behavior of every single propagator. Finally, the result will vanish if and only if every internal region is connected to exactly four other regions, i.e. for fishnet diagrams.

%%%%%%%%%%%%%%%%%%%%%%%%%%%%%%%%%%%%%%%%%%%%%%%%%%%%%%%%%%%%%%%%%%%%%%%%%%
\bibliographystyle{nb}
\bibliography{MassiveFishnets}

\end{document}